\documentclass[manuscript]{acmart}

\usepackage{amsmath, amssymb}

\usepackage[ruled,vlined]{algorithm2e}   

\usepackage{amsthm}
\usepackage{graphicx} 

\DontPrintSemicolon
\usepackage{tabularx} 
\usepackage{rotating}
\usepackage{booktabs}
\usepackage{subcaption} 
\usepackage[utf8]{inputenc}
\usepackage{longtable}
\usepackage{csquotes}
\usepackage{multirow}
\usepackage{enumitem}

\renewcommand\footnotetextcopyrightpermission[1]{} 
\usepackage{draftwatermark}
\SetWatermarkText{DRAFT – \today}
\SetWatermarkScale{0.4}
\SetWatermarkColor[gray]{0.98}

\title{Komitee Equal Shares: Choosing Together as Voters and as Groups with a Co-designed Virtual Budget Algorithm}

\author{Joshua C. Yang}
\email{joyang@ethz.ch}
\orcid{0009-0004-4044-6338}
\affiliation{%
  \institution{ETH Zurich}
  \city{Zurich}
  \country{Switzerland}
}

\author{Noemi Scheurer}
\email{noemi.scheurer@skkg.ch}
\orcid{0009-0006-4914-5747}
\affiliation{%
  \institution{Stiftung für Kunst, Kultur und Geschichte (SKKG)}
  \city{Winterthur}
  \country{Switzerland}
}

\begin{abstract}
Public funding processes demand fairness, learning, and outcomes that participants can understand. We introduce Komitee Equal Shares, a priceable virtual-budget allocation framework that integrates two signals: in voter mode, participants cast point votes; in evaluator mode, small groups assess proposals against collectively defined impact fields. The framework extends the Method of Equal Shares by translating both signals into virtual spending power and producing voting receipts. We deployed the framework in the 2025 Kultur Komitee in Winterthur, Switzerland. Our contributions are: (1) a clear separation of decision modes, addressing a gap in social choice that typically treats participatory budgeting as preference aggregation while citizens also see themselves as evaluators; and (2) the design of voting receipts that operationalise priceability into participant-facing explanations, making proportional allocations legible and traceable. The framework generalises to participatory grant-making and budgeting, offering a model where citizens act as voters and evaluators within one proportional, explainable allocation.

\end{abstract}

\begin{document}

\maketitle
\begin{figure}[h]
    \centering
    \begin{minipage}{0.47\textwidth}
        \centering
        \includegraphics[width=\linewidth]{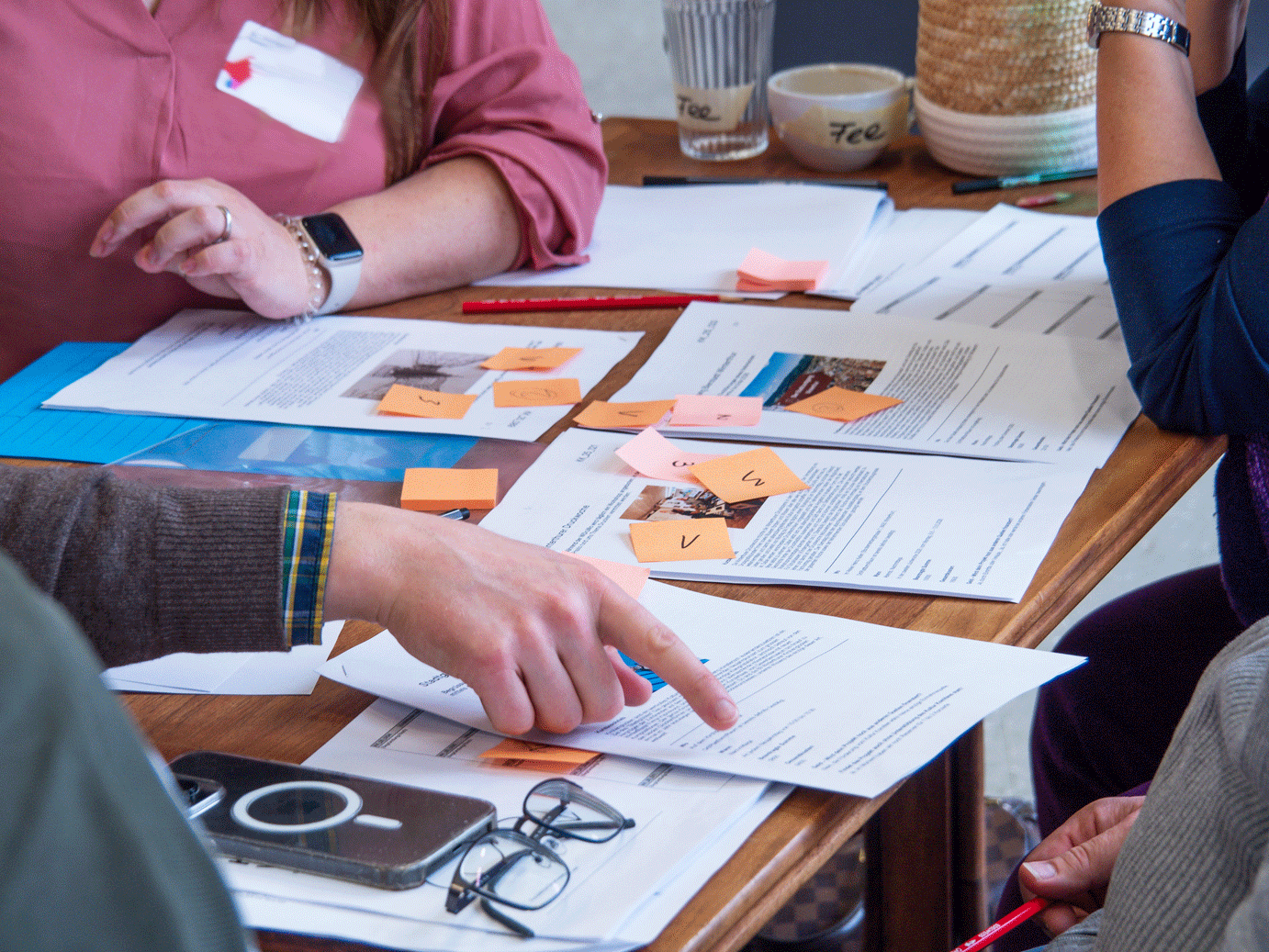}
    \end{minipage}
    \hfill
    \begin{minipage}{0.47\textwidth}
        \centering
        \includegraphics[width=\linewidth]{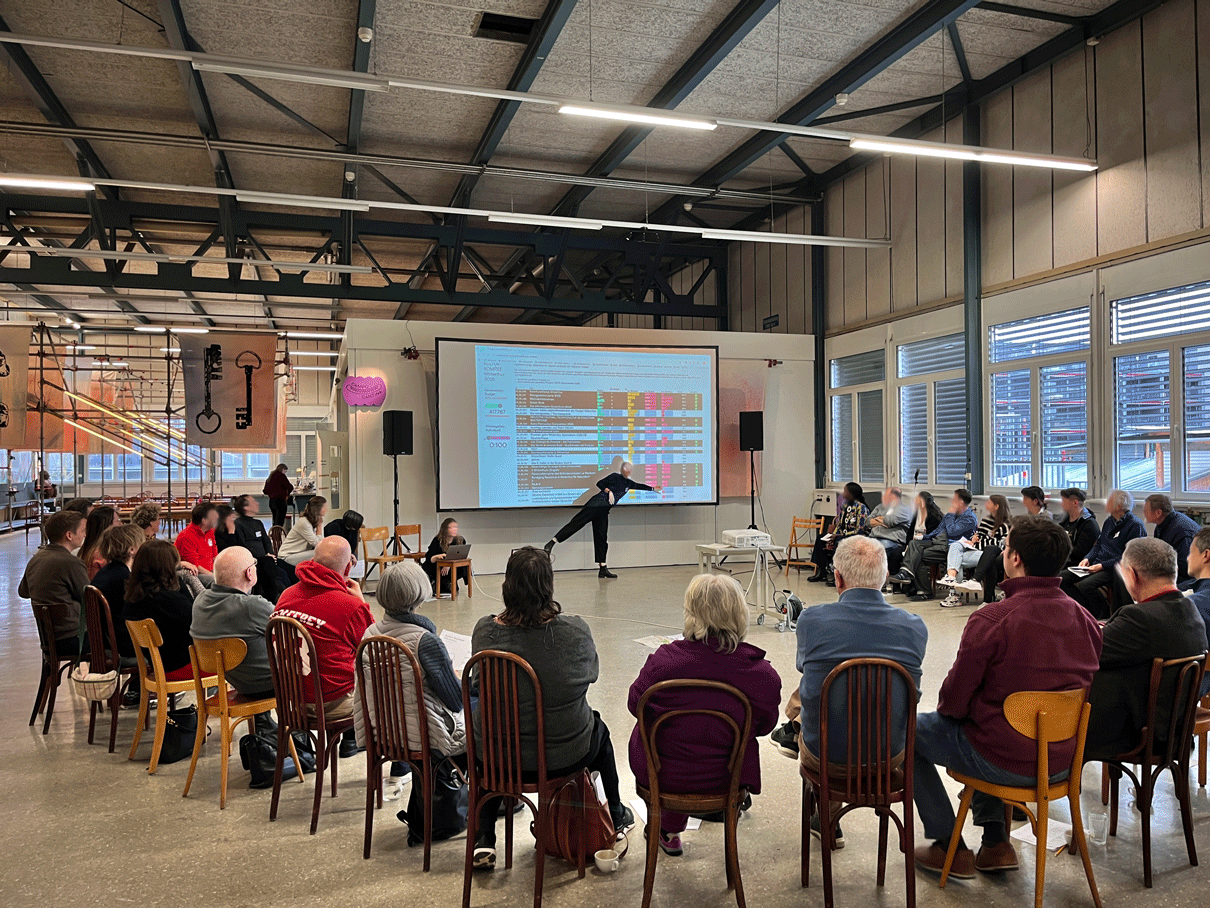}
    \end{minipage}
    \caption{Komitee Equal Shares process multiple real-world signals from individual voting and group deliberation into an aggregation algorithm that gives citizens a fair and explainable outcome.}
    \label{thumbnail}  
\end{figure}

\section{Introduction}

When citizens are asked to allocate shared funds, they face a dual challenge. Budgets are finite, project lists can run into the hundreds, and participants must weigh their own preferences against what they believe will serve the community as a whole. Processes that reduce this task to a simple tally of votes risk flattening civic judgement into private choice and leave participants wondering why certain projects succeeded while others failed.  

Participatory grant-making and participatory budgeting illustrate both the promise and the difficulty of such processes. Their appeal lies in giving ordinary residents a direct say in how public resources are distributed. Yet most implementations provide little support for the complexity of these decisions. Ballots often overwhelm participants, and outcomes are typically determined by greedy aggregation: projects are ranked by votes until the budget is spent. This logic privileges majority-backed proposals, sidelines minority priorities~\citep{Peters2020}, and ignores the fact that voters are not cost-conscious when confronted with dozens or even hundreds of proposals of varying sizes~\citep{Yang2024}. As a result, large projects may be funded disproportionately simply because they gather many votes, while smaller initiatives can be overlooked even when they would fit better within the budget.  

A deeper issue lies in how participation itself is understood. Social choice theory often treats a ballot as a proxy for personal preference. But in civic practice, participants frequently act with a broader mindset: they consider reach, equality, creativity, or tradition not only for themselves but on behalf of others. In other words, they are both voters and evaluators. Without clear roles or criteria, this dual responsibility becomes ambiguous. Citizens want to be fair judges for their community, yet lack guidance on whether to express private taste or public value. Research in democratic innovation has similarly noted that democratic participation is not only about casting votes, but also about how processes scaffold reflection, learning, and collective reasoning \citep{Chambers2023, Landemore2020, Hendriks2025}.  

The annual \emph{Kultur Komitee Winterthur} in Switzerland exemplifies this dilemma. Established in 2021 by the \emph{Stiftung für Kunst, Kultur und Geschichte} (Foundation for Art, Culture and History, SKKG), the Komitee invites around forty residents by sortition to distribute roughly CHF~400,000 each year to citizen-initiated cultural projects. Its goal is to diversify and democratise cultural funding by involving residents beyond established cultural stakeholders. Participants commit to several sessions in which they review applications, deliberate in groups, and make final funding decisions.  

\begin{figure}[h]
    \centering
    \includegraphics[width=0.8\linewidth]{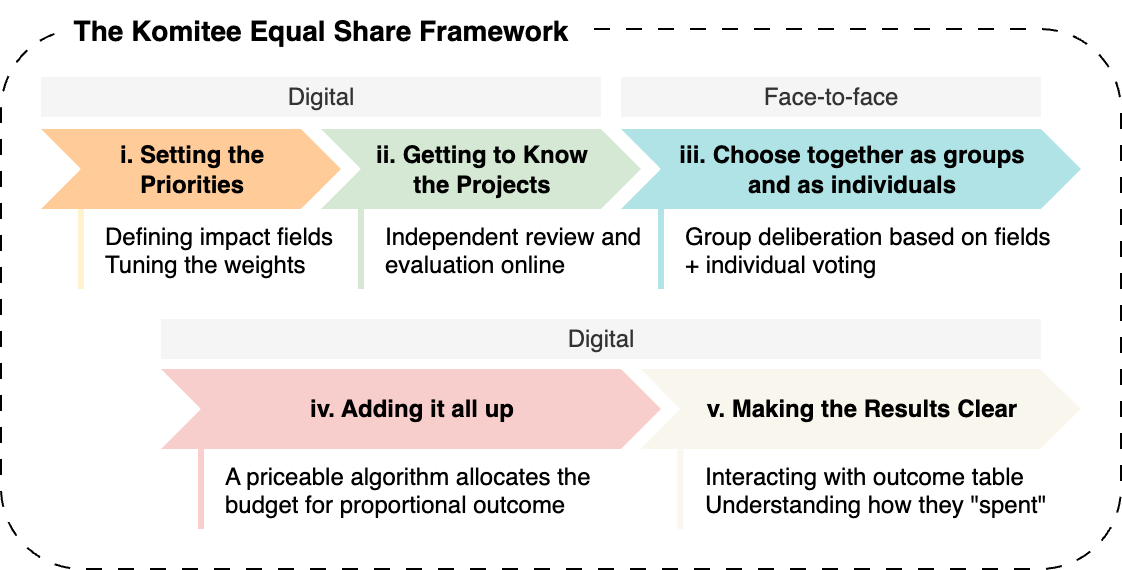}
    \caption{The Komitee Equal Shares framework as a five-stage process.}
    \label{fig:flow}
\end{figure}

This paper introduces \textbf{Komitee Equal Shares}, a framework and algorithm that combines the strengths of voting and deliberation into a single proportional allocation. Building on the Method of Equal Shares (MES) \citep{Peters2021}, it models participants in two explicit roles: as individuals with equal budgets (capturing private preferences) and as members of impact-field groups with weighted budgets (capturing civic evaluations). Projects are funded only when the combined virtual budgets of these agents can afford their cost, and each funded decision produces a \textbf{voting receipt} showing exactly how much support came from individuals versus impact fields. Priceability is significant here because it connects proportionality with explainability, making every outcome both fair and legible \citep{Rey2023, Peters2021}. Rather than treating justification as a post-hoc narrative, priceable allocations provide receipts as first-class artefacts that participants can inspect, reproduce, and contest. This resonates with work in HCI on procedural justice, which highlights how transparency and opportunities for outcome control can enhance perceived fairness and legitimacy in algorithmic mediation \citep{Lee2019}.  

We deployed Komitee Equal Shares in the 2025 Kultur Komitee (KK25), where 38 residents evaluated 121 proposals and allocated CHF~378,901 to 43 projects. Participants co-defined eight impact fields, weighted their relative importance online, and deliberated in two rounds of field-based groups before casting individual point votes. The algorithm integrated all signals into one unified allocation that participants could inspect immediately on decision day, supported by outcome tables and receipts.  

This study is guided by four research questions:  
\begin{itemize}
  \item[\textbf{RQ1}] How do different allocation rules shape which projects are funded?  
  \item[\textbf{RQ2}] How do participants perceive their own influence, and what budget split between individual and group decision-making do they view as fair?  
  \item[\textbf{RQ3}] How does the Equal Shares algorithm affect participants’ behaviour and their understanding of outcomes?  
  \item[\textbf{RQ4}] What do participants learn from the process, and how does it reshape their perspective on culture and the city?  
\end{itemize}  

In short, we examine allocation outcomes, perceptions of influence and fairness, behavioural adaptation to the algorithm, and the civic learning that participants gained through the process. By treating citizens simultaneously as voters, evaluators and co-designer of the algorithm, Komitee Equal Shares moves collective allocation beyond preference aggregation toward civic judgement. The framework merges the proportional guarantees of MES with HCI concerns of usability, transparency, and accountability. It provides a practical template for participatory grant-making: co-design impact fields, collect signals in both roles, run a priceable allocation, and publish voting receipts. In doing so, it enables communities to understand not just which projects were funded, but how and why those outcomes emerged from the combination of individual preferences and deliberative values.

\section{Background and Related Work}
\label{sec:background}

\subsection{Democratic Innovations beyond Voting}

Over the past three decades, a range of democratic innovations have sought to involve ordinary citizens directly in public decision-making. Two of the most prominent are participatory budgeting and citizen assemblies. Both move beyond traditional representative channels by giving lay citizens an explicit role in shaping collective outcomes, though they do so in different ways.  

Participatory budgeting (PB) is a process in which residents decide how to allocate a portion of public funds. Originating in Porto Alegre in 1989, PB typically invites project proposals from the public, organises deliberation in neighbourhood meetings, and concludes with citizen voting to determine which projects are funded. Early studies show that PB in Brazil not only broadened participation but also redirected resources toward poorer districts and basic services \citep{Baiocchi2005}. As PB spread to Europe, North America, and Africa, budgets were often smaller and participation lighter, yet PB continued to be promoted as a tool for legitimacy, inclusiveness, and trust in local government \citep{Sintomer2008,Wampler2012}.  

Citizen assemblies, by contrast, are deliberative “mini-publics” in which a randomly selected group of citizens, broadly representative of the population, deliberate on a policy issue and formulate collective recommendations. Supported by expert input, balanced briefing materials, and professional facilitation, assemblies provide ordinary citizens with the resources and time to deliberate productively on complex and divisive issues. Well-documented cases include the British Columbia Citizens’ Assembly on Electoral Reform (2004) \citep{Fournier2011,Smith2009} and the Irish Citizens’ Assembly (2016–2018), which shaped referenda on abortion and climate policy \citep{Farrell2019}.  

The significance of these innovations lies not only in participation but in the role of deliberation. Deliberative democratic theory has long argued that legitimacy depends less on aggregating fixed preferences than on the communicative processes that shape them. \citet{Habermas2015} stresses that democratic authority rests on inclusive and reasoned public deliberation, while \citet{Cohen1989} defines deliberative democracy as collective decision-making justified by arguments all can reasonably accept. \citet{Mansbridge1983} distinguishes adversarial models of democracy, which resolve conflict through voting, from unitary models that seek shared values through deliberation. Building on these foundations, \citet{Niemeyer2011} develops the concept of deliberative transformation, showing how deliberation can measurably reshape preferences and worldviews. Deliberative polling, pioneered by \citet{Fishkin2009}, shows that when citizens are provided with balanced information and structured dialogue, they often move from surface-level opinions to more considered judgements. Research on participatory budgeting and citizen assemblies similarly demonstrates that preferences are not fixed but reshaped through interaction. As \citet{Gutmann1998} argue, reason-giving transforms individual positions, while \citet{Dryzek2000} highlights how deliberation fosters recognition of others’ perspectives. \citet{Bachtiger2018} stress that conflict is not eliminated but channelled into structured exchanges, and \citet{Mansbridge2012} emphasise the systemic value of combining multiple sites of participation.   

Recent research shows that voting and deliberation are not alternatives but mutually reinforcing. \citet{Chambers2023} argue that voting strengthens deliberation by providing closure, ensuring equal status, structuring issues, and preserving dissent, while \citet{Hendriks2025} describe “participatory budgeting new style” as a hybrid model that combines large-scale online voting with deliberative forums. In a broader perspective, \citet{Smith2019} documents how hybrids connect assemblies to referenda and constitutional debates, and \citet{Landemore2020} portrays democracy as a system of practices in which citizens alternate between roles of voters, discussants, and evaluators. Such work highlights the rise of \emph{hybrid democratic designs}, which expand participation beyond casting votes or deliberating in isolation by weaving these practices together. The challenge is to design mechanisms of aggregation that respect this interdependence and enable citizens to act both as individual participants and as co-authors of collective decisions.

\subsection{Design Responses in Digital Participation}

Human–computer interaction research has increasingly addressed the challenge of structuring participation in ways that make aggregation and deliberation more transparent and inclusive. Early systems such as ConsiderIt \citep{Kriplean2012} demonstrated how lightweight interfaces could scaffold reflective public thought, while attention-mediation metrics \citep{Klein2012} showed how large-scale discussions might be made tractable. More recent work has examined how platform design shapes participation itself: online deliberation platforms create new opportunities for civic voice \citep{Davies2020}, and experiments with voting interfaces show that input design influences who participates, how clearly opinions are expressed, and whether outcomes are perceived as legitimate \citep{Benade2021,Hausladen2024}. Research on procedural justice further highlights that transparency and opportunities for outcome control are crucial for perceived fairness in algorithmic mediation \citep{Lee2019}. Input formats and interaction design are therefore not peripheral technicalities but central to the democratic character of digital systems.  

Beyond individual platforms, scholars have advanced sociotechnical frameworks that link democratic values to design practice. These frameworks emphasise involving underrepresented groups and evaluating systems not only by their technical performance but also by how well they support communication, inclusiveness, and deliberation \citep{Helbing2023,AbdelnourNocera2023}. Empirical studies show that digital tools can enable marginalised youth to express their views in public service design \citep{Dow2022}, facilitate deliberative talk in local governance \citep{Johnson2017}, and apply longitudinal analytics to monitor the quality of online engagement \citep{Shin2022}. Other research highlights the temporal dimension of participation: effective platforms must support not only one-off contributions but also sustained engagement over time \citep{Saad2018}. Experience-centred approaches extend this further, showing how participation can be embodied, aesthetic, and relational, challenging conventional boundaries between users, designers, and citizens \citep{McCarthyWright2024}.  

Recent innovations build on these foundations with computational support. \citet{Lu2025} present systems that scaffold civic capacities through structured prompts and background information, while \citet{Ma2025} propose methods for assessing the quality of human–AI deliberation. Algorithmic approaches are increasingly used to bridge voting and deliberation \citep{Yang2025}, and interactive visualisations help participants follow trade-offs and trace how their contributions are aggregated, thereby improving transparency and trust \citep{Kale2025}.  

This body of work shows that digital participation is not simply a technical problem but a sociotechnical design challenge. Interfaces, procedures, and evaluative frameworks shape whether citizens can inhabit their roles authentically, whether deliberation broadens perspectives, and whether collective outcomes are ultimately trusted.

\subsection{The Method of Equal Shares}

One of the more exciting recent developments in computational social choice is the \emph{Method of Equal Shares} (MES). Introduced by \citet{Peters2021} as a proportional voting aggregation method for participatory budgeting , MES has since moved beyond theory and into practical implementations. Its appeal lies in delivering \emph{proportional aggregation} with strong formal guarantees while remaining computationally manageable for real-world use.

MES performs \emph{proportional aggregation} by ensuring that every voter’s budget is gradually spent on projects they support, so cohesive groups secure proportional representation in the outcome. This contrasts with the greedy rules most commonly used in PB, where many votes are effectively ``unused'' once a project is funded. For example, if a project already wins under greedy, all additional approvals for that project are wasted, and smaller but cohesive groups may remain unrepresented. MES avoids this inefficiency by redistributing every voter’s share of the budget until it is fully exhausted.

Consider a set of voters $N = \{1, \ldots, n\}$ and a set of projects $P = \{p_1, \ldots, p_m\}$, each with cost $c_p > 0$. The total budget available is $B > 0$. Each voter $i \in N$ has an additive utility function $u_i : P \to \mathbb{R}_{\ge 0}$ that captures her preferences over projects, as defined in the previous section.

MES begins by assigning every voter the same \emph{virtual budget},  
\[
    b_i = \frac{B}{n} \quad \text{for each } i \in N,
\]
so that each voter starts with an equal share of the total budget.  

The key idea is that a project is fundable if its supporters can collectively cover its cost at a uniform ``price per unit of utility.'' For a non-negative parameter $q \ge 0$, a project $p$ with cost $c_p$ is said to be \emph{$q$-affordable} if
\[
    c_p \;=\; \sum_{i \in N} \min \bigl( q \cdot u_i(p),\, b_i \bigr).
\]
Here, $q \cdot u_i(p)$ is the amount voter $i$ would be willing to pay for project $p$ at price $q$ per unit of utility, and the minimum with $b_i$ ensures that no voter pays more than her remaining budget. If the sum of these capped contributions equals the project cost $c_p$, then the project can be purchased at price $q$.

The rule proceeds iteratively. It selects the project $p$ that becomes affordable at the lowest price $q$, adds it to the winning set, and deducts the contributions from each supporter’s budget:
\[
    b_i \;\gets\; b_i - \min(q \cdot u_i(p),\, b_i).
\]
Budgets are then updated, and the process repeats until no further project is affordable.

Over past few years, MES has been applied in PB programs in cities in Poland \citep{Boehmer2023} and Switzerland \citep{Pournaras2025}. Recent work has also built on MES to make it more practical in applied settings. The Adaptive Method of Equal Shares (AMES) \citep{Kraiczy2023} ensures that outcomes can be updated consistently when the available budget increases, without re-computation from scratch. The Bounded Overspending Equal Shares (BOS) rule \citep{Papasotiropoulos2025} allows limited budget overspending to fund broadly supported projects, reducing the number of voters left without funded representation. The Streamlining Equal Shares approach \citep{Kraiczy2025} introduces heuristics such as \emph{add-opt} to reduce unspent budget and demonstrates scalability on real-world PB instances. Empirical studies using open-source libraries such as Pabulib \citep{Faliszewski2023} further confirm that MES-like rules outperform greedy heuristics on fairness and robustness, while their polynomial-time implementation makes them tractable for practical use. Behavioural experiments also show that participants perceive MES as fairer and more legitimate than greedy allocation \citep{Yang2024}. 

The present paper builds directly on this line of work by introducing the \emph{Komitee Equal Shares} (KES) framework as an extension of MES. While MES treats each voter as an individual with an equal budget share, KES incorporates the additional layer of \emph{collective values} (impact fields) and deliberative inputs defined by a committee. In this sense, KES can be understood as a ``human wrapper'' around MES: it retains the proportional allocation logic of MES, but embeds it in a broader participatory process that combines individual voices, collective judgments, and hybrid online/offline interactions. This extension illustrates how the mathematical core of MES can be adapted to richer institutional designs for democratic decision-making.

\subsection{Priceability as Explainability}

A central property that underlies MES and related rules is \emph{priceability}. As \citet{Peters2020} first defined in the context of multi-winner voting, an allocation is priceable if voters could be imagined to hold equal amounts of virtual currency and, by following simple rules, to spend it only on projects they value in such a way that exactly the funded projects can be afforded. \citet{Peters2021} adapt this notion to participatory budgeting with cardinal ballots, showing that MES always produces priceable outcomes. Other rules also turn out to be priceable: sequential Phragmén and Maximin Support \citep{Brill2023} both generate outcomes that can be witnessed by a price system, and they connect directly to proportional justified representation.  

Priceability is significant because it connects proportionality with explainability \citep{Rey2023, Peters2021}. It guarantees that each voter begins with the same entitlement and that projects are only funded if their supporters can afford them collectively, ensuring proportional fairness. At the same time, priceability ensures that outcomes can be explained through a simple market analogy: citizens can reason about who “paid” for each project and why it was included. For social choice theorists, this property is valuable because it permits rigorous mathematical proofs of proportionality and fairness. Yet the real gap lies not in formal definitions but in translating them into forms that citizens of diverse educational backgrounds can understand. Bridging this gap requires work in human–computer interaction and civic design, where abstract guarantees must be turned into interfaces, metaphors, and feedback mechanisms that make sophisticated proportional aggregation legible and trustworthy in practice.  

Priceability provides a promising bridge, but it requires further HCI translation into terms and ideas that citizens can grasp. The challenge is therefore to design aggregation rules that make both roles explicit, that balance proportional fairness with explainability, and that generate outcomes which are not only correct in theory but also legible and meaningful in practice. It is precisely in this space that the Komitee Equal Shares framework is situated. By extending the priceable logic of MES to include both individual voting and group evaluation, it seeks to separate roles clearly, to generate ``voting receipts'' that show how projects were afforded, and to support fairness, learning, and legibility in participatory budgeting practice.

\section{Framework and Rationale}
\label{sec:framework}

Komitee Equal Shares is both a participatory framework and an allocation algorithm. The framework refers to the procedure around the rule: how citizens are convened, how they deliberate, and how their signals are collected. The algorithm is the computational core that aggregates these signals into a proportional and explainable allocation. In what follows, we first describe the framework as the participatory scaffolding that structures input into the process, and then turn to the algorithm itself, which extends the Method of Equal Shares (MES) to explicitly capture the dual roles of citizens as both voters and evaluators.

\subsection{The Komitee Equal Shares Framework}
\label{sec:framework}

The Komitee Equal Shares (KES) process unfolds in a series of steps that combine citizen participation with algorithmic allocation. Each step is designed to keep the procedure both fair and understandable.

\noindent\textbf{Step 1: Defining impact fields.}  
Before any projects are funded, the committee jointly develops and selects a small set of \emph{Wirkungsfelder}, or ``impact fields''. These fields articulate what the committee wants to make possible with its funding, such as inclusion, creativity, or support for newcomers. They are not exclusive criteria, but rather shared points of reference that make explicit the values against which projects will be judged. Because they are debated and revised by the committee itself, they encourage reflection on what the community cares about most.

\begin{quote}
\textbf{Note on terminology and algorithmic role.}  
The German term \emph{Wirkungsfelder} combines \emph{Wirkung} (effect, impact, influence) with \emph{Feld} (field, domain, sphere). More than just a set of abstract values or categories, it suggests domains where actions also generate tangible and observable real-world consequences. We translate it as “impact fields” to convey this sense of structured yet dynamic arenas of effect. Depending on context, we also use “values” to emphasise their normative orientation, or “criteria” to underline their evaluative function. Each translation highlights a slightly different dimension of the same idea: \emph{Wirkungsfelder} operate not only as standards for assessment but also as actionable reference points for shaping outcomes.

Within the Komitee Equal Shares algorithm, impact fields are translated into weights that tune the relative influence of deliberation versus individual votes. Each participant distributed points across the fields to indicate their importance. These distributions were aggregated into weights $w_v$, which endowed each ``value agent'' in the calculation with a proportional share of the collective budget. For example, if ``curiosity and interest'' received 19\% of points, the corresponding ``value agent'' in the algorithm gets allocated 19\% of the deliberative budget share. In this way, the committee’s collectively defined fields became algorithmic parameters that shaped which projects could be afforded and ultimately funded.
\end{quote}

\noindent\textbf{Step 2: Co-designing the weights of the impact fields.}  
Next, participants collectively decide how the algorithm should balance individual voices and shared values. Through an online vote, they determine (i) the fraction of the budget reserved for individual votes versus impact fields, and (ii) the relative weight of each field. In this way, the community itself co-designs the parameters of the allocation rule before it runs.

\noindent\textbf{Step 3: Online annotation homework.}  
Before the workshop, participants complete individual review tasks for all 30 project proposals. Each project is rated on a four-point scale (\emph{no, rather no, rather yes, yes}) and tagged with any number of impact fields. This stage is both evaluative and educational: by reading proposals and linking them to values, participants familiarise themselves with Winterthur’s cultural landscape and reflect on how community priorities apply in practice. The annotations provide baseline signals uninfluenced by group dynamics and structure the agenda for the subsequent workshop.

\noindent\textbf{Step 4: Deliberating in criteria groups.}  
In a face-to-face workshop, participants then rotate through small groups, each focused on one of the chosen criteria. Within each group, they discuss the projects and distribute points according to how well they think each project meets that value. The number of points a group has depends on the weight assigned to its criterion, so more important values carry more purchasing power later.

\begin{figure}[h]
    \centering
    \includegraphics[width=\linewidth]{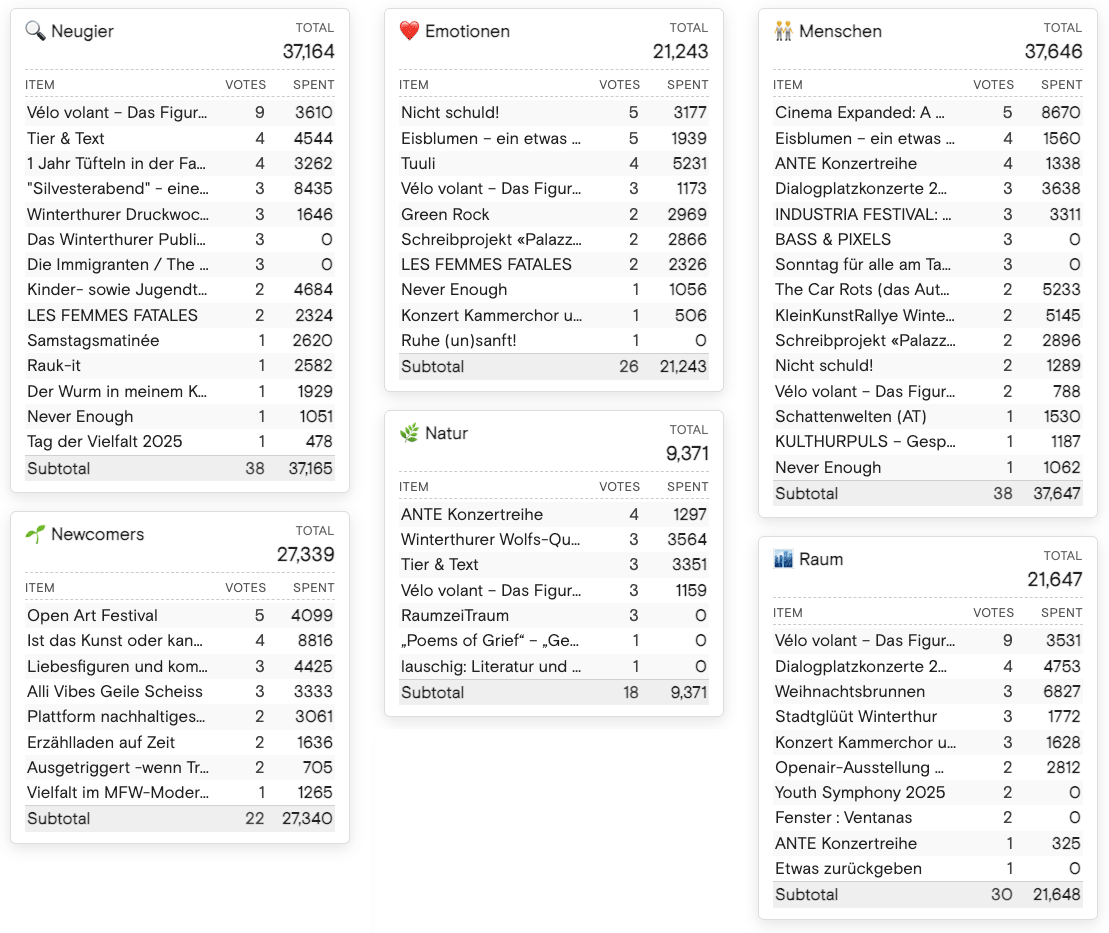}
    \caption{Screenshot of the voting receipts from the Komitee Equal Shares tool in Winterthur 2025. The receipts show how different groups and individual voters allocated their budgets across projects. This breakdown makes the algorithm’s decisions transparent by displaying both the votes and the actual spending for each group.}
    \label{fig:receipt}
\end{figure}

\begin{figure}[h]
    \centering
    \begin{minipage}{0.47\textwidth}
        \centering
        \includegraphics[width=\linewidth]{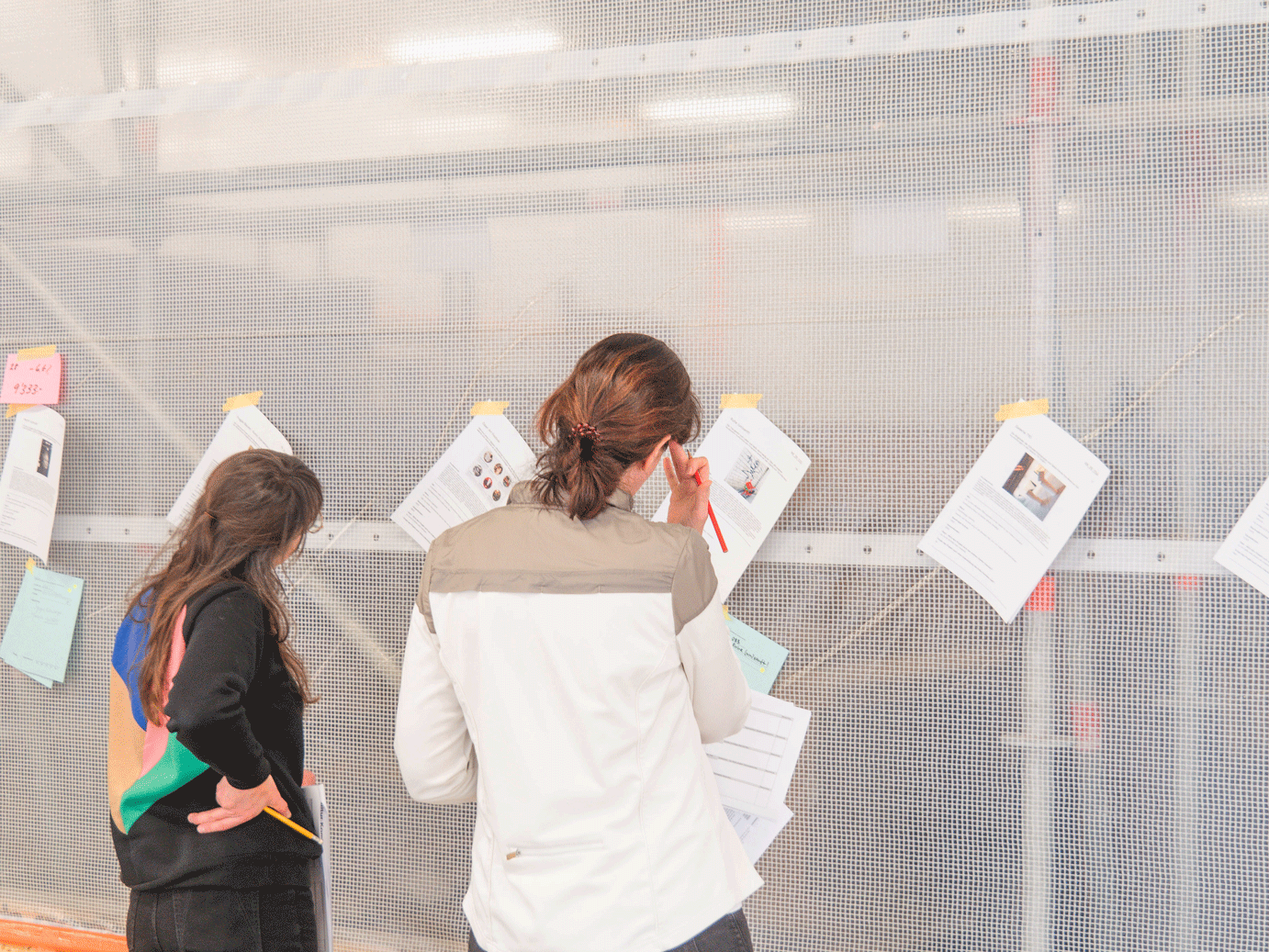}
        \caption{During the individual voting phase, the 121 projects and their proposals were placed around the venue for participants to walk around, explore, and contemplate which projects they should support.}
        \label{fig:indi_vote}
    \end{minipage}
    \hfill
    \begin{minipage}{0.47\textwidth}
        \centering
        \label{fig:outcome}
        \includegraphics[width=\linewidth]{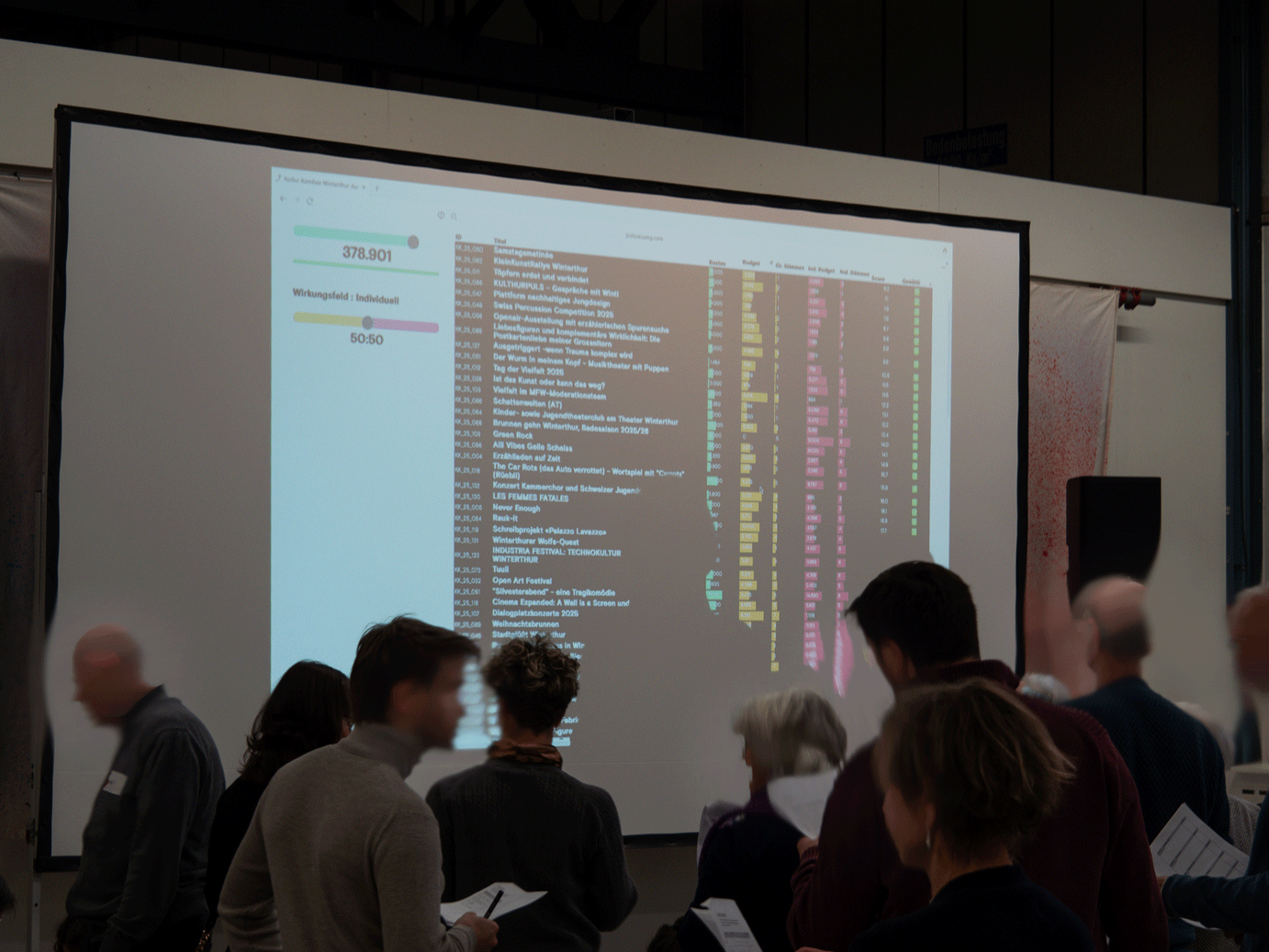}
        \caption{The outcome was calculated on the spot in real-time when all the individual votes and groups votes were collected and entered digitally. Participants could then interact with the result page and see how their votes contribute to the final outcome.}
        \label{fig:outcome}
    \end{minipage}
\end{figure}

\noindent\textbf{Step 5: Casting individual votes.}  
Alongside group deliberation, each participant also distributes a fixed number of points (for example, 20) across the projects they personally prefer. As Figure~\ref{fig:indi_vote} illustrates, participants walk around the room and identify the projects they wish to support. This individual vote captures sincere preferences, without requiring participants to strategise or rank-order everything. 

\noindent\textbf{Step 6: Splitting the budget and funding projects.}  
Once all signals are collected, the total budget is virtually divided into “buckets”: a share is split equally among individuals, while the remainder is allocated to values in proportion to their weights defined in Step 2. The algorithm then applies the logic of Equal Shares to fund projects one by one. A project is purchased if, and only if, its supporters—whether individuals or values—together have enough remaining budget to cover the cost. When a project is funded, the bill is split among its backers in proportion to their support. This algorithmic part of the process is described in detail in the following subsection (\S\ref{sec:equalshares}).

\noindent\textbf{Step 7: Exploring outcomes.}  
Participants can then view an interactive results table that shows which projects were funded and by whom (Figure~\ref{fig:outcome}). Sliders allow them to explore “what-if” scenarios, such as increasing the budget or adjusting the balance between individual and group influence. While these explorations do not change the official result unless formally agreed, they help participants see the trade-offs at stake.

\noindent\textbf{Step 8: Receiving voting receipts.}  
Finally, each participant and each value receives a voting receipt detailing how their virtual budget was spent. These receipts make the outcome legible: they show, for instance, that a theatre project was funded 60\% by individual enthusiasm and 40\% by the value of “supporting newcomers.” By closing the loop from preferences to projects, receipts provide accountability and transparency. Examples of these voting receipts are shown in Figure~\ref{fig:receipt}.

\subsection{The Equal Shares Algorithm}
\label{sec:equalshares}

\begin{algorithm}[H]
\caption{Core KES procedure (Equal Shares allocation with impact fields)}
\label{alg:kes_core}
\DontPrintSemicolon

\KwIn{$P$ projects with costs $C(p)$; total budget $B$; 
      individuals $\mathcal I$; 
      impact fields $\mathcal F$ with weights $w_f$; 
      utilities $u_a(p)\ge0$.}
\KwOut{Funded set $W$; receipts $\Delta_x(p)$ for each agent $x$.}

\BlankLine
\textbf{Preliminaries.}\;
Normalize utilities: $\hat u_a(p)\in[0,1]$ \tcp*{Scale utilities into [0,1]}
Assign each individual a bucket $b_i \leftarrow \tfrac{(1-r)B}{|\mathcal I|}$ \tcp*{Equal individual share}
Assign each impact field a bucket $b_f \leftarrow rB \cdot \tfrac{w_f}{\sum_{f'} w_{f'}}$ \tcp*{Weighted share for impact fields}
Set $W\leftarrow\emptyset$; \quad $\Delta_x(p)\leftarrow 0$ \tcp*{Initialize funded set and receipts}

\BlankLine
\While{true}{
  \ForEach{$p\in P\setminus W$}{
    $S(p)\leftarrow \sum_{i\in\mathcal I} b_i\,\hat u_i(p)$ \tcp*{Support from individual buckets}
  }
  $\mathcal P_{\mathrm{aff}}\leftarrow \{p: S(p)>0\ \land\ C(p)\le S(p)\}$ \tcp*{Projects affordable given support}
  \If{$\mathcal P_{\mathrm{aff}}=\emptyset$}{\textbf{break}} \tcp*{Stop if nothing affordable}

  pick $p^\star \in \arg\max_{p\in\mathcal P_{\mathrm{aff}}} S(p)$ \tcp*{Select project with strongest support}
  \ForEach{$i\in\mathcal I$}{
    $\Delta_i(p^\star)\leftarrow C(p^\star)\cdot \frac{b_i\hat u_i(p^\star)}{S(p^\star)}$ \tcp*{Split cost $\propto$ support}
    $b_i \leftarrow b_i-\Delta_i(p^\star)$ \tcp*{Update bucket}
  }
  $W\leftarrow W\cup\{p^\star\}$ \tcp*{Fund project}
}

\KwRet{$W,\Delta$} \tcp*{Final funded set and receipts}
\end{algorithm}

The Komitee Equal Shares algorithm, a part of the broader Komitee Equal Shares framework, builds directly on the \textit{Method of Equal Shares}, but extends it so that both individuals and collectively defined \emph{impact fields} act as agents with budget shares. Unlike the original method, where every agent has the same endowment, impact fields usually receive a larger share of the total budget, weighted according to their importance. In this sense, the procedure can be seen as a method of ``unequal shares,'' designed to balance personal preferences with community priorities. Conceptually, the allocation unfolds in five stages, which are mirrored in Algorithm~\ref{alg:kes_core}.

\noindent\textbf{Step 1: Giving every agent a budget.}  
Each individual receives the same virtual budget bucket, while each impact field (e.g., \emph{inclusion}, \emph{creativity}) is assigned a weighted bucket proportional to its priority. Together, these buckets define the resources available to fund projects (see \emph{Preliminaries} in the code).

\noindent\textbf{Step 2: Turning preferences into spending power.}  
When participants assign points to projects, these are normalised into support weights $\hat u_a(p)$. This ensures that each agent’s bucket is spent in proportion to their expressed support, avoiding distortions from differences in raw scoring scales (captured in the $S(p)$ calculation in Algorithm~\ref{alg:kes_core}).

\noindent\textbf{Step 3: Checking affordability.}  
A project can only be funded if the combined support of its backers is at least equal to its cost. The algorithm repeatedly selects the most strongly supported affordable project $p^\star$ (the “while affordable” loop).

\noindent\textbf{Step 4: Splitting the bill.}  
The cost of each funded project is divided among its supporters in proportion to their share of the support. This “priceable” mechanism ensures that no agent spends more than it has in its bucket (the update of $\Delta_x(p^\star)$ in the code).

\noindent\textbf{Step 5: Making the outcome legible.}  
Each funded project generates receipts showing exactly how costs were shared between individuals and impact fields. This provides a transparent audit trail ($\Delta$ returned at the end of the algorithm).

Komitee Equal Shares preserves the allocation logic of the Method of Equal Shares, but extends it with weighted impact fields and a surrounding deliberative process. Whether this still counts as strictly “proportional” in the social choice sense is open to debate: the unequal weights for impact fields, and the fact that priorities are often shaped through real human deliberation, mean that proportionality here is not purely axiomatic but socially constructed. What the system does guarantee is that no agent—whether individual or impact field—ever spends more than its assigned budget, and that every funded project is fully paid for by its backers. The outcome is therefore not only feasible and transparent, but also grounded in a hybrid of algorithmic fairness and human judgment.

\medskip

\subsection{Why this design matters}

At the algorithmic core lies the \textit{Method of Equal Shares}, a proportional allocation rule where every agent receives the same budget share and can only fund projects up to that limit. Komitee Equal Shares extends this logic in a crucial way: it introduces \emph{impact field agents}—collectively defined groups such as \emph{inclusion}, \emph{creativity}, or \emph{heritage}—with larger buckets to spend. Instead of every agent holding an equal share, impact fields receive budgets in proportion to the weights assigned by the committee. This small change has a large effect: it opens a space for deliberation about which impact fields matter, how much they should be resourced, and how they ought to shape the portfolio of projects. 

In this sense, Komitee Equal Shares is not only an algorithm but a framework. The skeleton remains Equal Shares, but it is wrapped in a process design that combines digital and face-to-face participation, voting and deliberation, individual preferences and community-defined \emph{impact fields}. The design deliberately bridges traditions from participatory budgeting and citizen assemblies, and connects computational social choice with HCI concerns such as usability, transparency, and accountability. 

The weighting of impact fields and the influence of real-world deliberation mean that proportionality here is partly socially constructed rather than purely formal. What is guaranteed is that no agent—individual or impact field—ever spends more than its assigned budget, and every funded project is fully covered by its backers.

This layered design yields several properties:

\begin{itemize}
  \item \textbf{Coalitional fairness.}  
  As in Equal Shares, any coalition—whether individual participants or impact field agents—can only fund up to its budget. This prevents domination and ensures that support is always tied to resources.  

  \item \textbf{Dual representation.}  
  Participants express personal preferences, while impact fields are represented as weighted agents. Both are integrated in the same allocation, making visible the interplay between individual enjoyment and shared principles.  

  \item \textbf{Unified allocation.}  
  Instead of splitting the budget into separate pots, Komitee Equal Shares blends signals into a single allocation. Projects tend to succeed when individual preference and impact field support align, producing hybrid outcomes by design.

  \item \textbf{Explainable transparency.}  
  Every funded project generates a receipt that shows the exact contributions of individuals and impact fields. These receipts act as explanation artefacts, providing a legible ledger of why each project succeeded and how resources were combined.  
\end{itemize}

In short, Komitee Equal Shares allows citizens not only to express preferences but also to deliberate about the collective criteria that deserve spending power, weaving together algorithmic guarantees with human judgment.

\section{Case Study: Kultur Komitee Winterthur}

The \emph{Kultur Komitee} (Culture Committee) in Winterthur is a democratic innovation that redistributes cultural funding through randomly selected citizen deliberation. Established in 2021 by the \emph{Stiftung für Kunst, Kultur und Geschichte} (Foundation for Art, Culture and History, SKKG), the project channels part of the foundation’s endowment back into the local cultural landscape. Each year, about 200 residents are drawn by lot from the city’s population register, of whom roughly 10–15\% accept the invitation, resulting in a committee of thirty to forty members who allocate CHF~400,000 (approx.\ USD~450,000) to cultural projects. While the Komitee does not claim to be demographically representative, its main goal is to diversify and democratise art and culture funding by bringing in voices beyond the usual cultural stakeholders. Participants commit to several sessions where they learn about the city’s cultural landscape, review applications, deliberate in groups, and make final funding decisions.

We describe the Komitee as a \emph{Budget Assembly}: a hybrid of participatory budgeting and citizen assemblies applied here to cultural grantmaking, but adaptable to other domains of public funding. Participants are not only asked to vote for the projects they personally prefer, but also to reflect on shared values on behalf of the wider community. Transparency and traceability are explicit goals: the process is designed to be both fair and understandable, so that participants and the public can see how funding decisions were made and what priorities they embody.

\subsection{Learning from 2024: Split Voting and Deliberation}

In 2024, the Kultur Komitee tested a two-phase hybrid process. In the first phase, members voted online, and half of the budget (CHF~190,000) was allocated using the \emph{Method of Equal Shares}, funding 17 projects. At the workshop, participants had the chance to review this algorithmic outcome, made only one small budget adjustment, and then approved the rest.  

In the second phase, the remaining CHF~190,000 was distributed through group deliberation. Participants ranked projects in small groups, which resulted in 18 additional projects being funded. In total, 35 projects received support. Importantly, the two phases rested on different logics: voting aggregated individual preferences, while deliberation fostered community-oriented reasoning that only emerged in discussion.  

This design surfaced two key lessons. First, because voting and deliberation covered different subsets of projects, the most popular proposals were never scrutinised collectively, while deliberation was concentrated on borderline cases. The result was an uneven use of collective intelligence, applying individual judgment to some projects and community judgment to others. Second, several participants reported uncertainty about their role: should they support projects aligned with personal taste and hobbies, or should they judge on behalf of the city as a whole? While they previously developed a set of joint criteria, this \emph{role ambiguity} left some unsure whether they were “doing their job correctly,” and highlighted the need for clearer guidance on how individual and collective responsibilities should be balanced.

\begin{figure}[ht]
    \centering
    \begin{minipage}{0.48\linewidth}
        \centering
        \includegraphics[width=\linewidth]{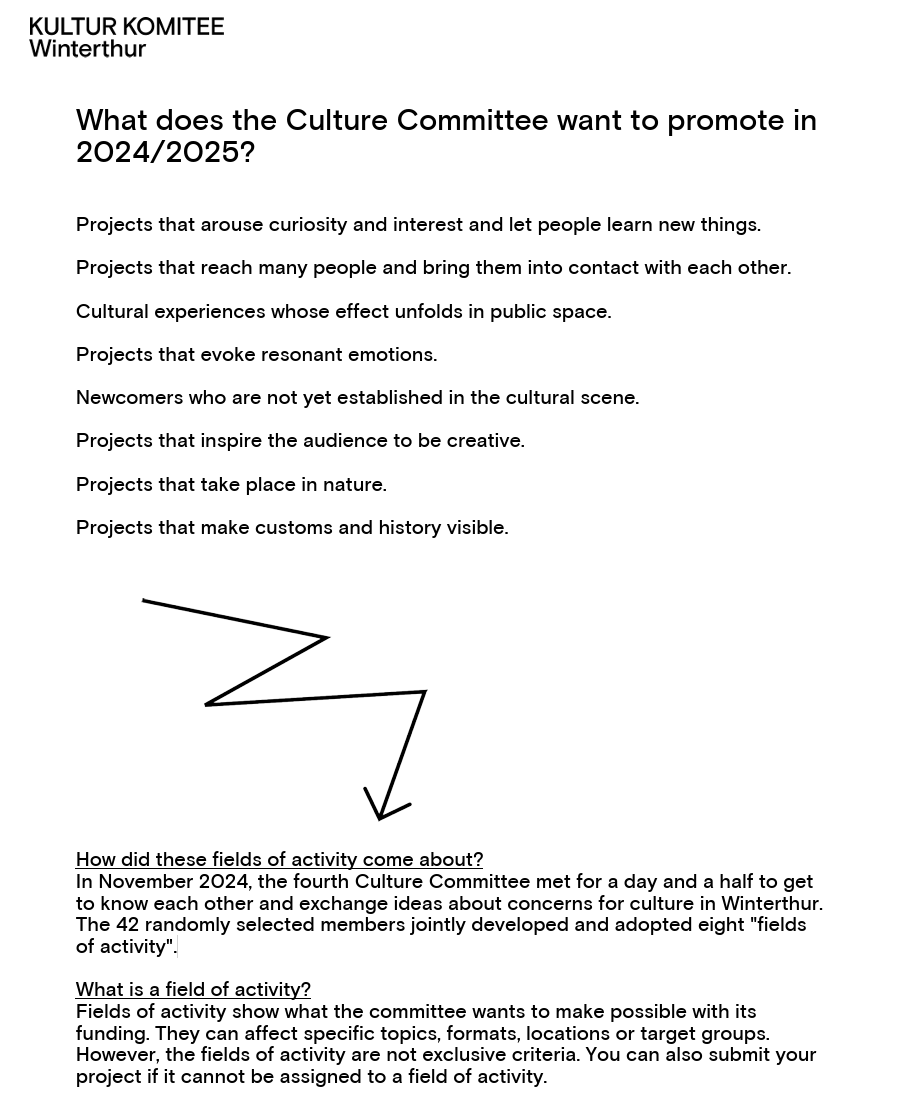}
        \caption{The screenshot of the official call for cultural projects in Winterthur in November 2024. The public document includes the eight jointly defined fields of activity of the 2025 Kultur Komitee in Winterthur, which guided the funding priorities and were later translated into weighted value categories.}
        \label{fig:fields_activity}
    \end{minipage}
    \hfill
    \begin{minipage}{0.50\linewidth}
        \centering
        \begin{minipage}{0.48\linewidth}
            \centering
            \includegraphics[width=\linewidth]{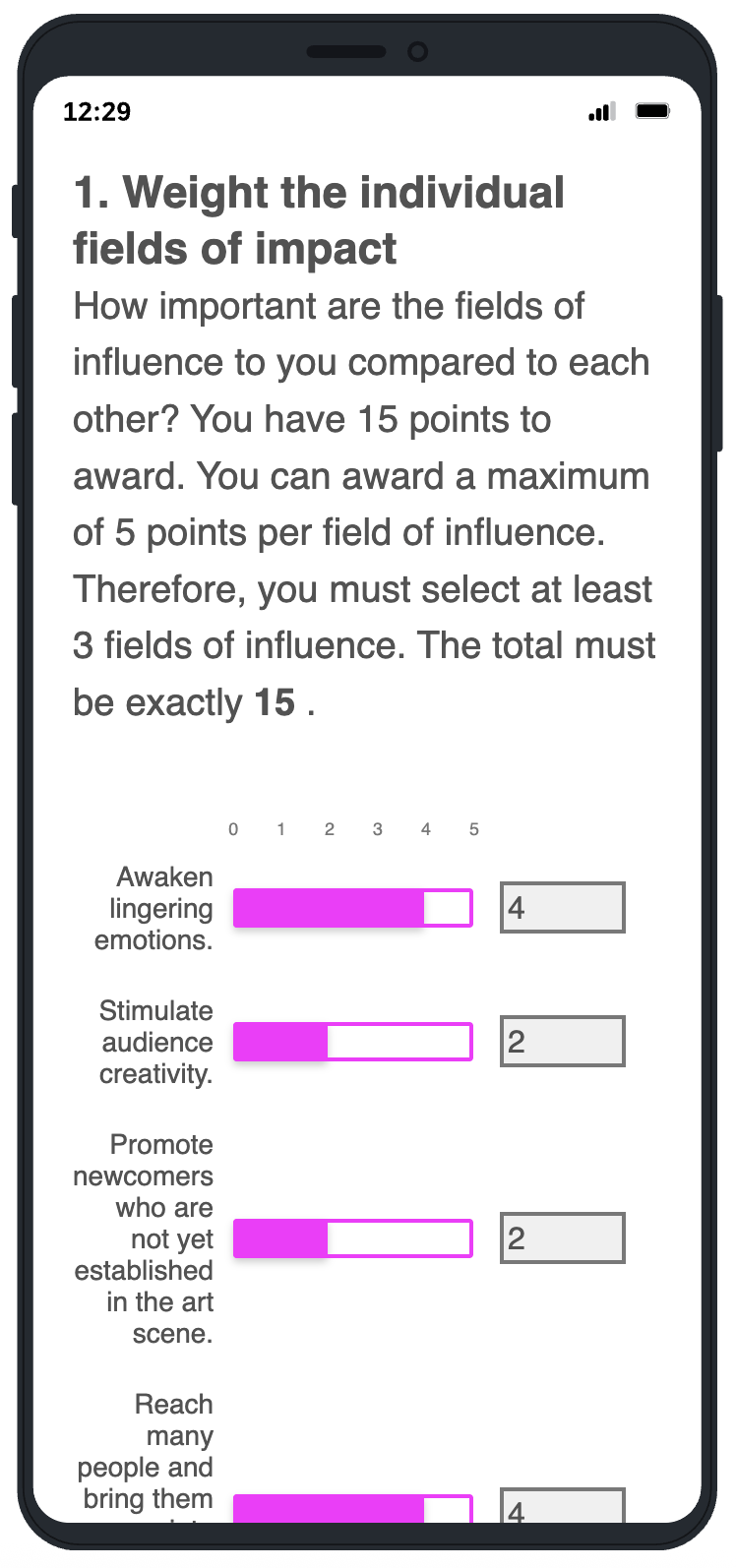}
        \end{minipage}
        \hfill
        \begin{minipage}{0.48\linewidth}
            \centering
            \includegraphics[width=\linewidth]{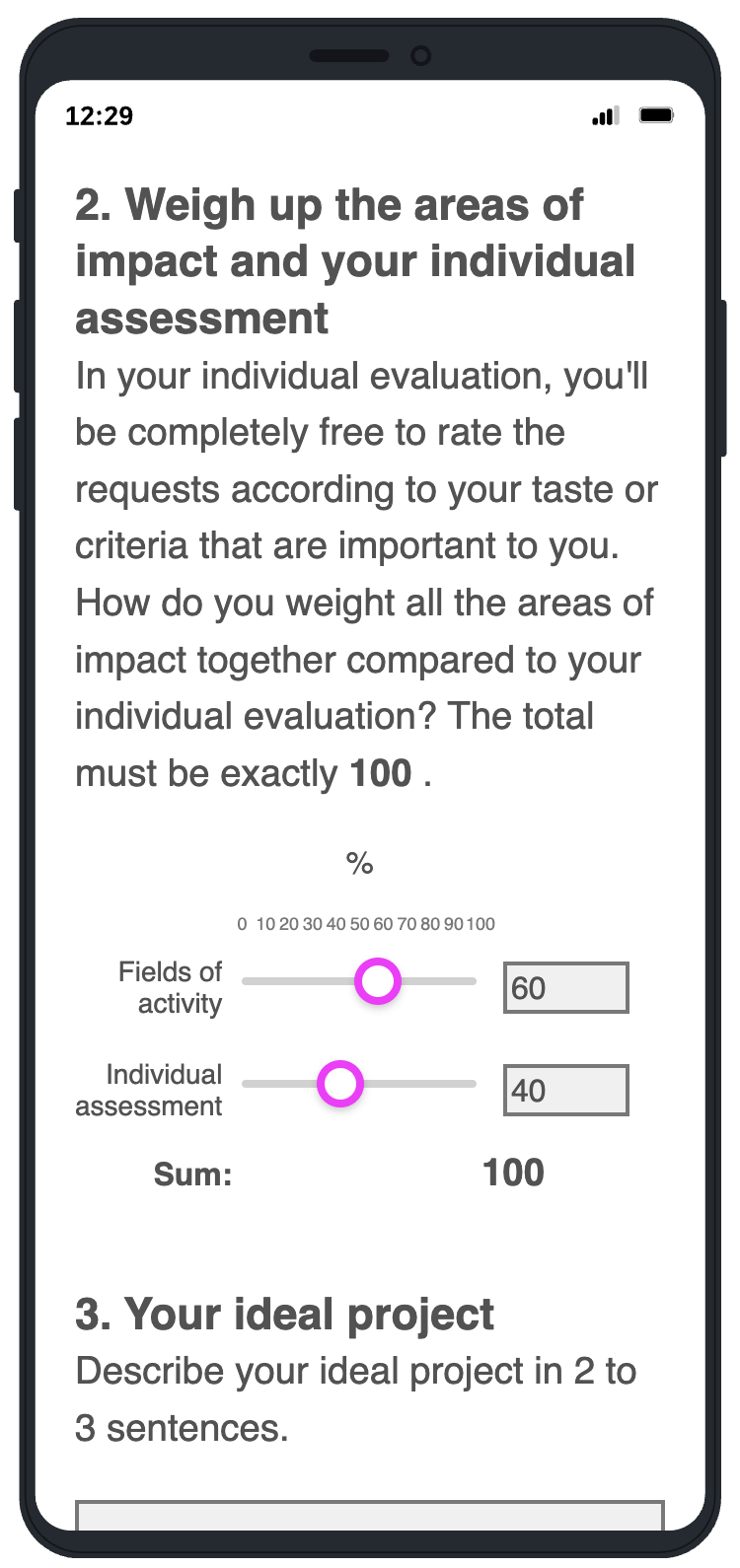}
        \end{minipage}
        \caption{Participants first distribute 15 points across fields of cultural impact, fine-tuning the relative importance of each criterion within the Komitee Equal Shares algorithm. They then set the balance between these collectively defined field weights and their own individual assessments, deciding how much each mode of evaluation contributes to the final outcome.}
        \label{fig:parameter_tuning}
    \end{minipage}
\end{figure}

\subsection{A Unified, Impact-Field Allocation in 2025}

In November 2024, the fourth Kultur Komitee met in person for a day and a half to get to know each other and discuss cultural priorities in Winterthur. During this workshop, the 38 randomly selected members jointly developed and adopted eight \emph{fields of impact}. These fields articulated what the committee wanted to make possible with its resources—shared priorities rather than rigid eligibility criteria—and were further refined with guiding questions and potential indicators.  

After the workshop, members participated in an online exercise to quantify these priorities, as illustrated in Figure~\ref{fig:parameter_tuning}. Each participant distributed points across the eight fields to indicate their relative importance, producing weights that translated the qualitative discussions into numerical parameters for allocation. In the same exercise, members also decided how the total budget should be split between collective signals (impact fields) and individual signals (personal votes). Preferences varied widely—from as little as 20\% to as much as 80\% for the collective share—but the distribution centred closely around parity (mean = 49.5\%, median = 50.0\%, SD = 14.5\%). So the split was decided to be 50:50. 

The final weights of the eight fields of impact were: Curiosity \& Learning (19\%), Community \& Connection (18\%), Public Space \& Accessibility (15\%), Emotional Impact (13\%), Support for Newcomers (11\%), Audience Creativity (10\%), Nature \& Environment (9\%), and Tradition \& History (6\%). These eight fields are then published in public in the official call for projects (Figure \ref{fig:fields_activity}), clearly announcing to the project proposers that these are the impact fields that the projects would be evaluated on. 

Within the Komitee Equal Shares framework, each field was represented as an \emph{impact-field agent} with a budget proportional to its assigned weight. As described in the steps of previous Section~\ref{sec:framework}, this stage functioned as participatory co-design. Participants not only cast votes but also set the weights of impact fields and the balance between collective and individual signals, thereby shaping the allocation rule itself.

\begin{figure}[t]
    \centering
    \includegraphics[width=1\linewidth]{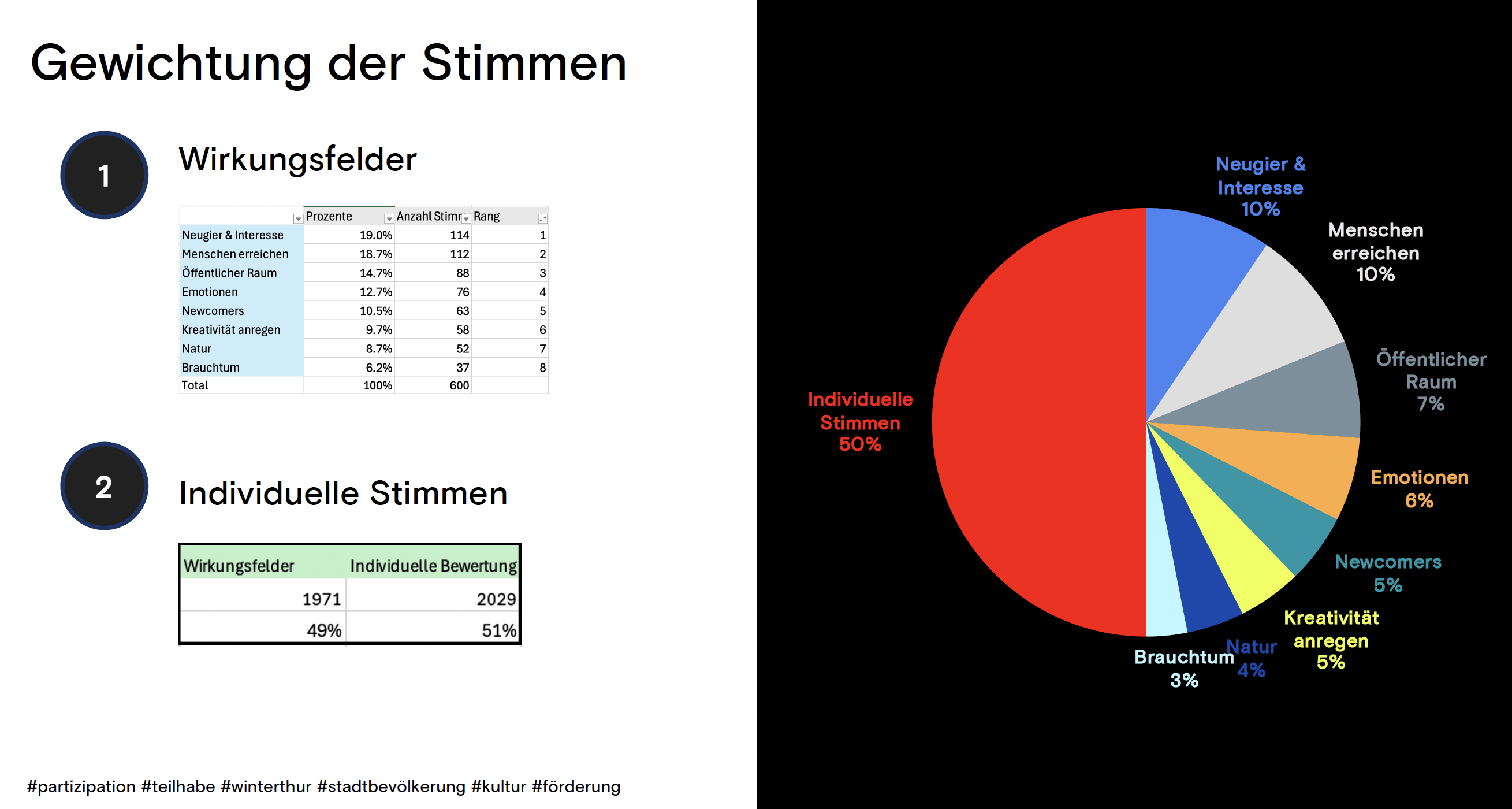}
    \caption{The Explaination Shown to the 2025 Kultur Komitee during the Workshop. Left: impact fields with their assigned percentages and corresponding group sizes/points; right: overall budget split between individual votes (50\%) and impact fields (50\%), showing the proportional contribution of each field.}
    \label{fig:weights}
\end{figure}

The decision day followed a carefully structured format. In the morning, participants joined focus groups dedicated to the eight impact fields. Each person took part in two rounds, meaning that every participant contributed to the deliberation of two different fields. Within these groups they first reviewed the assigned projects, then discussed them collectively, and finally distributed group points.  

Both the size of the groups and the number of points available were scaled in proportion to the weight of each impact field (Figure~\ref{fig:weights}). While such weights could have been multiplied only later in the algorithm, we deliberately applied them directly to the group design. This made the final outcome table easier to understand: project scores simply added up individual and group points on the same scale. Because group points were already weighted, the total score could be read as a measure of combined support without hidden multipliers. This satisfied participants’ natural desire to see which projects were most popular and avoided the difficulty of interpreting “mixed” signals in raw Equal Shares outputs.  

In the afternoon, participants entered an individual phase. Each person received 20 points to distribute across at least three projects of their choice. Around 4pm, after the individual ballots were completed, both the group results and the individual votes were entered digitally. The data was formatted in a ``.pb'' file (commonly used for Participatory Budgeting aggregation) and processed through the Komitee Equal Shares tool, which automatically generated the final allocation. Because the procedure was largely automated, the outcome could be displayed almost immediately after the last votes were cast.  

Participants then turned their attention to the outcome table. The main result was projected onto a large screen so that everyone could see which projects had been funded. At the same time, participants were able to browse the table, interact with the page, and explore the details of the allocation. The hosts explained the results, highlighted which projects had succeeded, and demonstrated the voting receipts that showed how individual votes and impact-field budgets had contributed. This collective reveal in the physical space ensured that it remains a moment of common discussion and interpretation (Figure~\ref{fig:outcome}).

After the allocation was finalised, each participant was invited to choose one or two project each to act as \emph{Götti/Gotti} (mentors). This role did not affect the allocation itself but created a sense of follow-up responsibility: over the following year, Göttis/Gotti were expected to check in on their adopted projects and report back, strengthening the connection between the KK participants and the cultural initiatives they had helped fund.  

At the end of the day, the Komitee Equal Shares algorithm integrated both signals into a single proportional allocation of the CHF~380,000 budget. Projects could request between CHF~3,000 and CHF~40,000, and the Komitee was entitled to reduce requests by up to 20\%. This rule was frequently applied to balance fairness with budget discipline. Proposals came from both professional and amateur cultural actors, requiring citizens to weigh institutional projects alongside grassroots initiatives. From the perspective of each participant, half of their virtual budget counted through their individual vote, and the other half was expressed through the impact-field agents, ensuring that the final portfolio reflected both personal preferences and collectively defined cultural priorities.

\begin{figure}
    \centering
    \includegraphics[width=\linewidth]{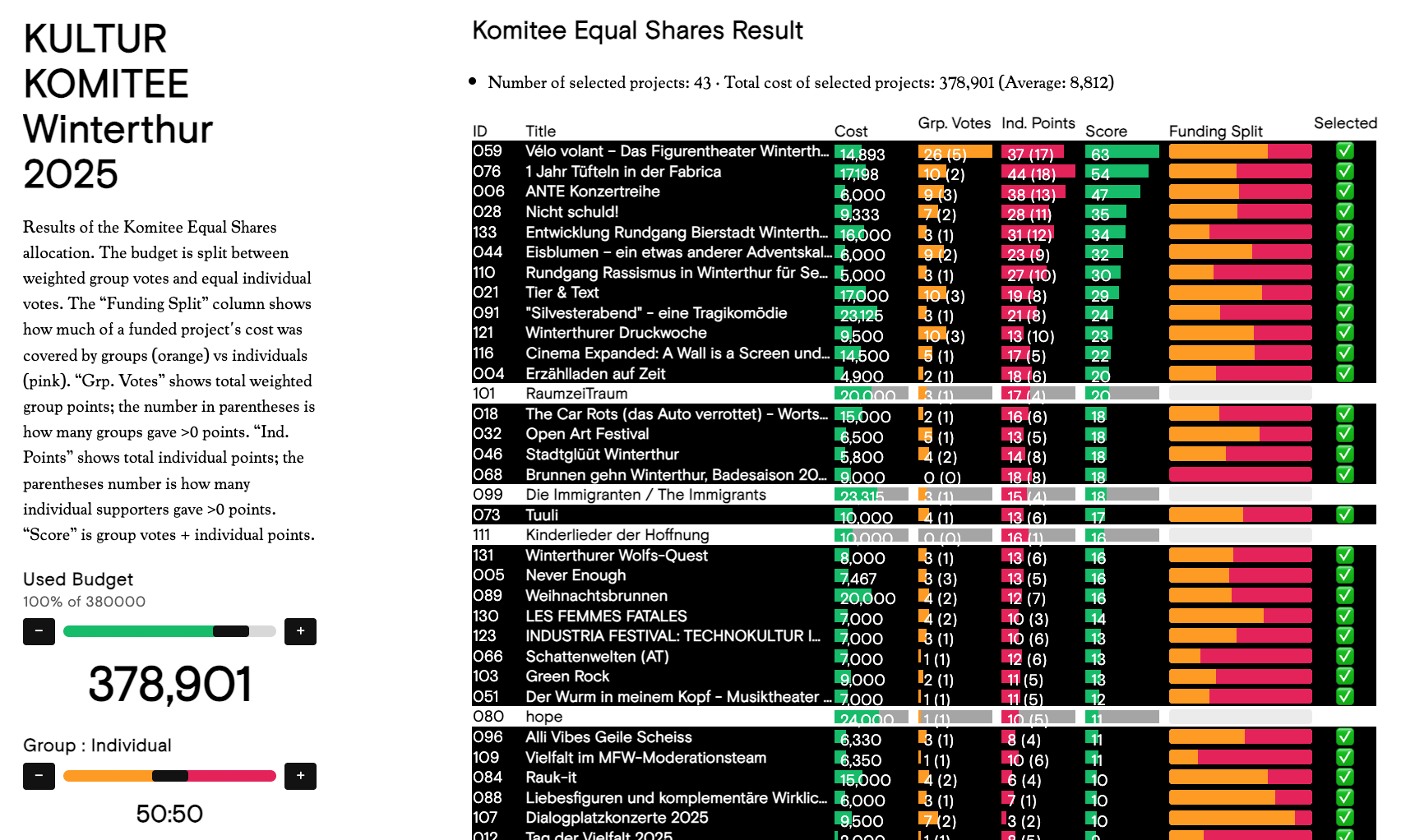}
    \caption{Screenshot of the Komitee Equal Shares (KES) tool used in Winterthur 2025. 
The interface visualises how the community budget was distributed across projects, 
combining weighted group votes and equal individual votes. The results show the 
selected projects, their costs, and how funding was split between collective group 
decisions and individual support, producing a fair and explainable allocation outcome. The score column, simply an aggregation of all the points assigned, is added for providing an intuitive understanding of the level of support a project has. The original version shown to Winterthur residents was in German.}
    \label{fig:kes_screenshot}
\end{figure}

\paragraph{Example.} The project \emph{Vélo volant – Das Figurentheater Winterthur radelt ins Quartier} (CHF~14,893) illustrates how the dual signals worked in practice. It received 37 points from 17 individual voters and 26 points from value-based deliberation. Neither channel alone would have secured funding; combined, they were sufficient. The case shows how projects needed to appeal both to individual participants and to the collectively defined impact fields in order to succeed.

\section{Method}

\subsection{Methodological Approach}
This research is situated within traditions of \emph{participatory design} and \emph{field deployment}. Rather than testing an artefact in a controlled setting, we co-designed and deployed a democratic innovation ``in the wild'' in a participatory grantmaking endeavour in Winterthur, Switzerland. The process was shaped collaboratively: the lead author engaged in regular planning meetings with the organising team, and one organiser is a co-author of this paper. The design was iteratively adapted in response to citizen feedback during earlier rounds, consistent with participatory and co-design practices in HCI. This project exemplifies collaboration between \emph{researchers and practitioners}, with academic inquiry embedded in a live civic process.

\subsection{Participants and Procedure}
Initially, in October 2025 a total of 250 residents of legal age were drawn by lot from the municipal regestry. Of those contacted, 42 Winterthur residents agreed to join the \emph{Kultur Komitee}. During the early process, four individuals decided to drop out due to personal circumstances, which resulted in a final commitee of \textbf{38  members}. 

The participants committed to join at least four in person meetings from as well as taking part in online activities from November 2024 till April 2025:
\begin{enumerate}
    \item \textbf{Kick-off and Workshop : }Onboarding and outlook on the process \textbf{
    \item Workshop on Decision Making: }Joint exchange and development of decision making criteria (impact fields) 
    \item \textbf{Open Call for projects: }Winterthur residents and instituions submit project proposals
    \item \textbf{Online voting}: Comitee members review around 30 projects and cast an initial vote from home
    \item \textbf{Decision making workshop: }Collective decision making based on joint criteria as well as individual votes\textbf{ (Deliberation points}, allocated in structured group workshops to reflect collective judgments on community values;)
    5. \textbf{Exchange with funded projects: }Completion of the funding process

\end{enumerate}
\begin{table}[h]
\centering
\caption{Demographic composition of the 2025 Culture Committee.}
\label{tab:demographics}
\begin{tabular}{ll}
\toprule
\textbf{Category} & \textbf{Distribution} \\
\midrule
Gender & Male: 20 (51.3\%), Female: 19 (48.7\%) \\
Age & Median: 52; Min: 25; Max: 79; Std: 14.2 \\
     & 18--30: 3 (7.7\%), 31--40: 9 (23.1\%), 41--50: 8 (20.5\%), \\
     & 51--60: 3 (7.7\%), 61--70: 10 (25.6\%), 71--80: 5 (12.8\%) \\
Education & Tertiary: 24 (61.5\%), Secondary II: 13 (33.3\%), \\
          & Compulsory school: 1 (2.6\%) \\
Origin & Switzerland: 31 (79.5\%), Germany: 2 (5.1\%), Spain: 2 (5.1\%), \\
       & Kenya: 1 (2.6\%), Australia: 1 (2.6\%), \\
       & Turkey: 1 (2.6\%), China: 1 (2.6\%) \\
\bottomrule
\end{tabular}
\end{table}

To acknowledge the substantial time and cognitive effort required, all participants received \textbf{CHF~500 in compensation}, covering both their online and in-person workshop activities. In addition, a supplementary fund was established to support participants with special needs, such as care responsibilities for relatives or loss of income due to their involvement. The composition of the KK 2025 committee is described in Table \ref{tab:demographics}. It is important to emphasise that the sortition procedure is always followed by a stage of self-selection. Because participation in the committee is voluntary, the resulting composition reflects broader patterns of volunteering in Switzerland. For example, older citizens tend to volunteer more frequently than younger ones, and higher levels of education are generally associated with higher levels of volunteering. As a result, the committee is not fully representative of the population of Winterthur. We acknowledge that this demographic composition also influenced the design of the hybrid process, which was tailored to the needs and capacities of the participants. 


\subsection{Data Sources}

Our analysis draws on three complementary types of data produced at different stages of the decision-making process:
\begin{itemize}
    \item \textbf{Pre-workshop annotations.}  
    In advance of the workshop, all 38 committee members independently reviewed the 30 submitted project proposals. For each project, they provided a four-point rating (\emph{no, rather no, rather yes, yes}) and tagged it with any number of relevant impact fields from the set of criteria jointly defined by the committee. These annotations form the baseline of individual preferences and value attributions.

    \item \textbf{Workshop deliberation and voting.}  
    The in-person workshop was held on March 1, 2025, with 36 participants attending. In the morning, participants engaged in two rounds of small-group deliberation, each table centred on a specific impact field. Groups collectively distributed a fixed number of “deliberation points” to projects they deemed best-suited for the impact field. In the afternoon, all participants cast their own individual votes across projects. Both the deliberation points and the individual votes serve as core inputs to the \emph{Komitee Equal Shares} algorithm, jointly determining the funded portfolio under the shared budget.

    \item \textbf{Post-workshop survey.}  
    Following the workshop, 32 of the 36 participants completed an online survey. The instrument combined Likert-scale items with open-ended questions, capturing motivations, experiences, and evaluations of the process. This survey data complements the allocation records by providing insight into participant reasoning and perceptions of fairness. Translated to English using DeepL.
\end{itemize}

Together, these sources provide a \emph{mixed-methods evaluation}, combining behavioural traces, collective allocations, and self-reported reflections.

\subsection{Ethical Considerations}
Participation in the \emph{Kultur Komitee} is voluntary and governed by a research agreement that participants sign at the outset. The agreement clarifies that data may be used for research purposes, but only in anonymised form. Researchers access the data under a contractual arrangement with the organising team, ensuring that only de-identified, secondary data are transferred and analysed. The study protocol received approval from the lead author’s research institution. These measures align with established ethical standards for human-centred computing research.

\subsection{Researcher Positionality}
The lead author participated in the co-design of the process alongside the organising team, and one practitioner from the organising team is also a co-author of this paper. This dual role as both researcher and co-designer carries the risk of bias but also ensures a deep, situated understanding of the process. We acknowledge that this study is not a detached observation but rather a \emph{collaboration between researchers and practitioners} aimed at advancing both practical democratic innovation and theoretical understanding in HCI.

\section{Results}

\subsection{Determinants of Final Individual Votes}

The \emph{Komitee Equal Shares} process generated multiple signals before participants cast their final individual votes. In the pre-deliberation stage, participants rated around thirty projects in an online survey and described their ``ideal'' project. While these online ratings did not formally count toward allocation, they were used to structure the workshop by clustering projects into thematic groups. They thus provide a baseline measure of individual preferences. 

During the workshop, participants engaged in two rounds of small-group deliberation within their assigned field groups, where projects were discussed and points were allocated. In the plenary stage, participants were also shown the aggregate support that other groups had given to projects before making their final individual allocations in the afternoon.

Table~\ref{tab:final_vote_predictors} reports the logistic regression results in the form of odds multipliers (OR) per one-standard-deviation increase. An OR greater than~1 indicates a higher likelihood of giving points; an OR less than~1 indicates a reduced likelihood. Asterisks denote predictors significant at $p < .001$.

\begin{table}[ht]
\centering
\caption{Predictors of giving points in the final individual vote. Odds multipliers (OR) per +1 SD, with 95\% confidence intervals. Predictors are grouped by stage of the process.}
\label{tab:final_vote_predictors}
\begin{tabular}{lccc}
\toprule
Predictor & Stage of Process & OR (95\% CI) & Significance \\
\midrule
Online rating of projects& Pre-deliberation & 1.69 [1.54–1.86] & * \\
Full-text similarity (ideal vs. project) & Pre-deliberation & 1.12 [0.98–1.28] & \\
Group deliberation – Round 1 (own group) & Group deliberation & 1.70 [1.50–1.92] & * \\
Group deliberation – Round 2 (own group) & Group deliberation & 1.86 [1.67–2.08] & * \\
Broad group support (other groups) & Collective signals (plenary) & 2.94 [2.23–3.86] & * \\
Unexpected group support (residual) & Collective signals (plenary) & 0.38 [0.28–0.53] & * \\
Project budget (log) & Project characteristic & 1.03 [0.90–1.19] & \\
\bottomrule
\end{tabular}
\end{table}

The results suggest a layered decision process. Pre-deliberation signals acted as an anchor: pre-deliberative online ratings (OR 1.69) and a trace of semantic similarity between their self-described ideal projects and the project description (OR 1.12, not statistically significant) carried through to the final stage. Although weaker than deliberative effects, these anchors show that participants’ personal preferences continued to shape their decisions rather than being completely overwritten by group influence.  

Deliberative exposure had the strongest impact. Projects discussed within a participant’s own groups were much more likely to receive support (ORs 1.70 and 1.86). Even more influential was broad endorsement from other groups (OR 2.94). This can be explained in several ways. Participants may have felt they had already supported projects in their two assigned fields and used their individual votes to balance their influence by recognising projects from other areas. At the same time, endorsements from other groups provided an informational shortcut: they signalled peer validation and highlighted projects with a higher chance of success under the Equal Shares allocation. Broad group support therefore guided attention, encouraged balancing behaviour, and reinforced strategic reasoning.  

Interestingly, a corrective dynamic was also evident. Projects that received more group support than expected, given their online ratings, were actually less likely to gain additional individual points (OR 0.38). This suggests that participants sometimes held back when a project already appeared to have sufficient momentum to succeed, choosing instead to allocate their limited support to other proposals still in need of votes.

Finally, project cost had no effect (OR 1.03). Participants supported inexpensive and expensive projects alike, potentially because they trusted the Komitee Equal Shares algorithm to handle budgetary constraints. This allowed them to focus on the cultural and social value of projects rather than financial calculations.  

Overall, participants acted as their own aggregators—anchoring decisions in personal preferences, integrating deliberative exposure, responding to collective signals, and strategically redistributing support.

\subsection{Budget Distribution Outcome}  

\subsubsection{Different Funding Scenarios}

\begin{table}[h]
\centering
\caption{Comparison of portfolios generated under three scenarios.}
\label{tab:portfolio_comparison}
\begin{tabular}{lccc}
\toprule
Scenario & Projects funded & Total budget & Avg. budget \\
\midrule
Individual-only (greedy) & 35 & 379,925 & 10,855 \\
Group-only (greedy)      & 33 & 378,712 & 11,476 \\
Komitee Equal Shares     & 43 & 378,901 & \textbf{8,812} \\
\bottomrule
\end{tabular}
\end{table}

For comparison, we constructed two hypothetical portfolios using the aggregation logic most common in participatory budgeting. In the \emph{group-only} scenario, only the deliberative point allocations were counted, and projects were funded greedily from highest to lowest until the CHF~380,000 budget was exhausted. In the \emph{individual-only} scenario, the same procedure was applied using only individual votes. These serve as baseline conditions because in most PB practice, outcomes are determined by simple vote counts and greedy spending without any proportional rule. The actual funded portfolio, by contrast, was produced by the \emph{Komitee Equal Shares} method, which integrates both group and individual signals within a priceable, cost-conscious allocation rule.

\begin{figure}[h]
    \centering
    \includegraphics[width=0.7\linewidth]{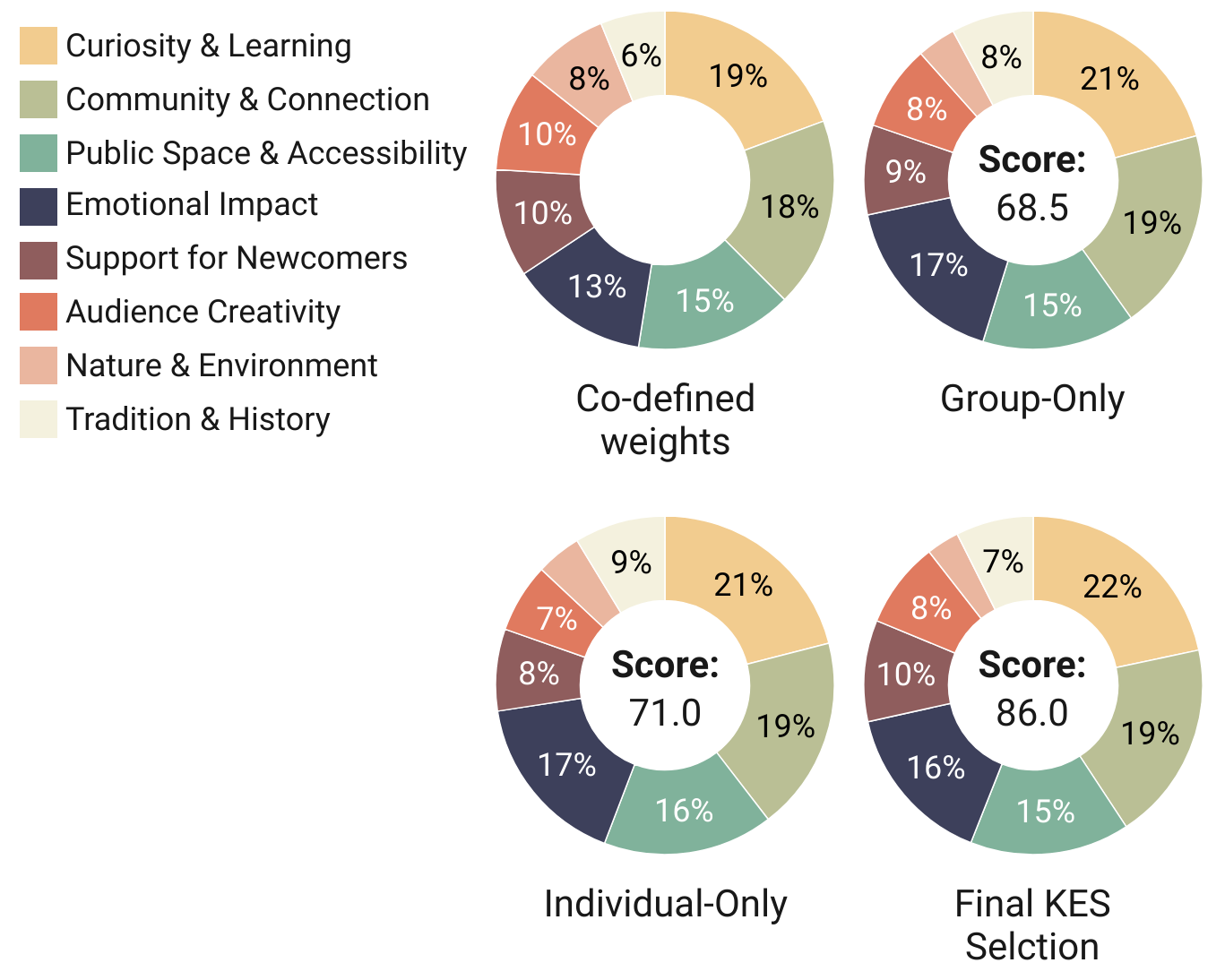}
    \caption{Comparison of value alignment across funded portfolios. Each ring shows the proportional representation of the eight co-defined criteria in different scenarios. From left to right, top to bottom: (1) the initial co-defined weights set by the committee, (2) a portfolio selected using only group deliberation votes, (3) a portfolio selected using only individual votes, and (4) the final allocation under \emph{Komitee Equal Shares}. Scores in the centre represent the overall aggregate fulfilment of community-defined values, based on pre-workshop project annotations. }
    \label{fig:alignment}
\end{figure}

The three scenarios produce notably different portfolios. The individual-only and group-only baselines fund around the same number of projects (35 and 33, respectively), with average project budgets exceeding CHF~10,000. By contrast, the final portfolio produced through \emph{Komitee Equal Shares} supports a broader set of 43 projects, with a markedly lower average budget per project (CHF~8,812). This reflects a shift toward greater diversity in the final outcome: more projects receive support under the same budget constraint, especially smaller initiatives that might otherwise have been crowded out by larger proposals.

This pattern connects directly to the aforementioned OR results in table~\ref{tab:final_vote_predictors}. Participants did not systematically reward or penalise projects based on their budget size—cost was not a significant predictor of votes. Yet Equal Shares, as a priceable allocation method, inherently incorporates project budgets: a project can only be funded if its accumulated support from groups and individuals is sufficient to cover its cost. In practice, this means participants could focus on the content and value of projects, while the method itself ensured budget feasibility. The result is a portfolio that balances citizen preferences with affordability, funding a larger and more diverse set of projects without exceeding the total cap.

\subsubsection{Alignment with Impact Fields}
To assess whether the final portfolio of funded projects reflected the community’s impact fields, we compared it with the two counterfactual scenarios. Prior to the deliberation workshop, each participant read a random sample of 30 project proposals online and annotated them with any number of the eight co-defined impact fields. This annotation process produced a grounded value profile for every project, which we then used to compute the aggregate impact fields distribution of different funded portfolios.  

Figure~\ref{fig:alignment} visualises these comparisons through the lens of criteria. The leftmost ring shows the \emph{co-defined weights}, i.e., the relative importance of the eight criteria as set collectively by the committee at the start. The second ring shows the \emph{group-only} portfolio: its value distribution closely mirrors the predefined weights, reflecting the proportional design of the deliberative point allocation. The third ring, the \emph{individual-only} portfolio, also tracks the predefined weights closely, since individual ratings were informed by and guided by the group process.  

The rightmost ring displays the \emph{final portfolio} under Komitee Equal Shares. Here, the distribution remains well aligned with the predefined criteria, but the total score is substantially higher (86.0 compared to 71.0 and 68.5). This increase is a direct consequence of the Equal Shares method: because it is cost-conscious, it avoids over-investing in a few expensive projects and instead funds a larger number of smaller projects that still fit the community’s values. The result is both a balanced profile across criteria and a higher aggregate fulfilment of the values the committee had set out at the beginning.

\subsection{Perceived Influence and Preferred Budget Splits} 

\begin{figure}[ht] 
\centering 
\includegraphics[width=0.6\linewidth]{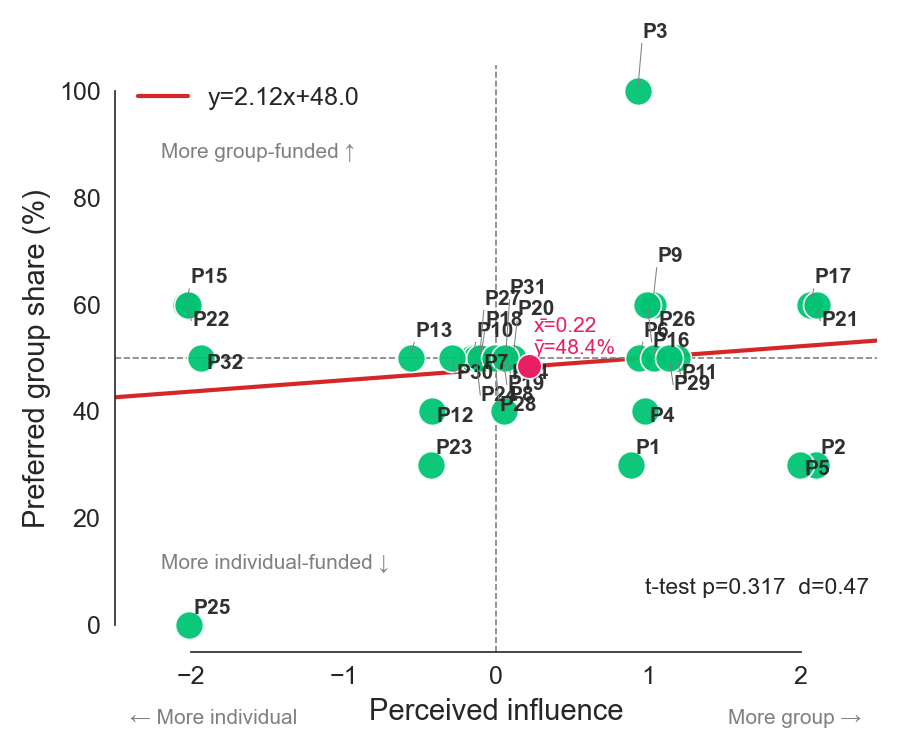} 
\caption{Scatter plot of perceived self-influence (x-axis) versus preferred budget allocation to group deliberation (y-axis). While outliers exist, most participants converge around a 50:50 split.} 
\label{fig:influence_share} 
\end{figure} 

A central theme in our study is how participants experienced their own role in the dual channels of decision-making: as individual voters and as members of deliberating groups. To capture this, we asked them (1) how they perceived their own influence in the final outcome, and (2) how they would ideally design the system in terms of the budget split between individual and group decision-making. 

Figure~\ref{fig:influence_share} compares these perceptions and preferences. The horizontal axis encodes perceived self-influence (negative = individuals felt stronger, positive = groups felt stronger, zero = equal). The vertical axis shows participants’ normative preferences for how much of the budget should be decided in group deliberation. Each circle represents a participant (labelled by ID). The regression line indicates the best linear fit; the magenta point marks the sample mean ($\bar{x}=0.22$, $\bar{y}=48.4\%$). 

Overall, the relationship was weak and non-significant ($t=0.317$, $d=0.47$). Crucially, across the spectrum of perceived self-influence, most participants clustered around a 50:50 allocation. This suggests that even when individuals sensed asymmetries in how much their own voice mattered, they nevertheless endorsed parity between individual and group channels as the fairest design. Rather than adjusting the rules to amplify their own channel of influence, most accepted the principle of equal weighting between individual and group decision-making, seeing it as a just balance that could accommodate differing perceptions of power.

Survey results reinforce this interpretation. When asked about their satisfaction with participation opportunities in the selection of applications, 22 respondents (68.8\%) reported being “satisfied” and 8 respondents (25.0\%) “rather satisfied,” meaning that nearly 94\% expressed positive views. Only 2 respondents (6.2\%) were “rather dissatisfied.” This distribution suggests that participants not only converged on a balanced design normatively, but also felt they had meaningful opportunities to influence the outcome in practice.

\begin{table}[h]
\centering
\caption{Participant reflections on influence and fairness of the budget split.}
\label{tab:influence_quotes}
\begin{tabular}{p{1.2cm}p{12cm}}
\toprule
\textbf{ID} & \textbf{Quote} \\
\midrule
P9  & ``Good mix of different decision-making phases.'' \\
P17 & ``Ultimately, projects that receive the broadest recognition from the group should be funded. This was the case. The decision was developed and evaluated collectively.'' \\
P29 & ``I was clearly in the minority with my priorities. But I'm quite satisfied; my ego isn't so attached to the final decision :-).'' \\
P15 & ``Everyone had full opportunity to express their opinion at all times. There's nothing to criticize.'' \\
P7 & ``I had the feeling that the individual voices were somewhat lost in comparison to the weight of the fields of influence.'' \\
P21 & ``The committee members had the opportunity to carefully review and evaluate the submitted projects and actively influence the funding decisions through their votes. Furthermore, constructive exchange within the Cultural Committee was encouraged, allowing different perspectives to be considered. The fact that the projects I voted for received financial support demonstrates that the participation process was effective and taken seriously.''\\
P25 & ``I think the method is good. People who aren't familiar with the culture would probably be overwhelmed if they had to decide everything on individual votes. I did it like I do in politics: Distribute all the votes among a few (3) projects, confident that other good projects would receive votes from others.''\\
P29 & ``Everyone was given sufficient space, often and intensively, to contribute their decisions.'' \\
\bottomrule
\end{tabular}
\end{table}

These reflections reinforce that the perceived fairness of the 50:50 split was not only statistical, but also grounded in participants’ lived sense of voice, inclusion, and procedural trust. P25’s comment illustrates how the dual-role design worked in practice: the group phase acted as scaffolding, easing the cognitive burden for those less familiar with the cultural landscape, while the individual phase enabled strategic expression of personal priorities. By concentrating votes “as in politics” yet trusting others to support remaining projects, P25 shows how deliberation and individual choice complemented each other, with each role preparing and legitimising the other.

\subsection{How the Equal Shares Algorithm Shaped Decisions}

\begin{table}[ht!]
\centering
\caption{Participant reflections on how the Komitee Equal Shares algorithm influenced their decisions.}
\label{tab:es_influence}
\begin{tabular}{p{1.2cm}p{2.5cm}p{10cm}}
\toprule
\textbf{ID} & \textbf{Influence} & \textbf{Quote} \\
\midrule
P2   & Yes (strategic) & ``No longer supporting my 'heart projects' with individual votes.'' \\
P8   & Yes (fairness)  & ``A little. In the individual voting, I eliminated all the projects that had already received very high group votes.'' \\
P12  & Yes (relief)    & ``It took a lot of pressure off of having to do everything 'right and fairly.' That was great.'' \\
P16  & Yes (relief)   & ``It was a relief for me to know that such a complex key was used and that this resulted in a great deal of justice.'' \\
P21  & Yes (strategic) & ``I supported projects that reflected both my views and broad support within the committee. I also paid attention to realistic budget requirements to promote a fair distribution of resources.'' \\
P26  & Yes (strategic) & ``Sure, I didn’t even consider projects without opportunities.'' \\
P31  & Yes (adaptive)  & ``There was a tendency within the groups to support more, cheaper projects rather than a few, expensive ones. I adapted to this.'' \\
\midrule
P7   & Mixed (reassured) & ``It reassured me that balance was ultimately maintained.'' \\
P11  & Mixed (uncertain) & ``No, I didn't quite understand the method—but it was good to know that somehow everything was taken into account. :-)'' \\
P25  & Mixed (unaware)   & ``I somehow wasn't aware of that. I think the method is good... I did it like I do in politics: distribute all the votes among a few projects.'' \\
\midrule
P27  & No & ``No, this did not affect me.'' \\
P32  & No (but fair) & ``No: And I think it's fair the way they did it.'' \\
\bottomrule
\end{tabular}
\end{table}

Beyond overall perceptions of fairness, we also asked participants directly whether the use of the Komitee Equal Shares algorithm influenced their own decisions. Responses reveal a wide spectrum: some participants adjusted their strategy in light of the algorithm, while others felt it did not affect their behaviour at all.

As detailed in Table~\ref{tab:es_influence}, out of 32 respondents, around one third (10 participants, 31\%) reported that the algorithm influenced their decisions. Some described how the algorithm relieved pressure and guided their voting strategy. A further 7 participants (22\%) gave mixed or uncertain responses. Some expressed reassurance without clear behavioural change, while others admitted confusion or lack of awareness. The remaining 15 participants (47\%) reported no influence. Some were brief and unequivocal, while others emphasised that they accepted the fairness of the outcome even if it did not affect their behaviour. The implication is that proportional algorithms should be able to operate effectively even when not all participants internalise their mechanics- as trust in the principle of balance can be as important as detailed understanding.

\subsection{Explainability Under Real-World Constraints}

\begin{table}[h]
\centering
\caption{Participant reflections on explainability, categorised by sentiment.}
\label{tab:explainability_quotes}
\begin{tabular}{p{0.8cm}p{1.5cm}p{12cm}}
\toprule
\textbf{ID} & \textbf{Sentiment} & \textbf{Quote} \\
\midrule
P8  & Positive & ``The forms (result table) also seemed clear to me.'' \\
P12 & Positive & ``In general, I found all the information clear, simply structured, and reliable.'' \\
P18 & Positive & ``I was 100\% confident that this was a well-thought-out and effective approach, and I'm overall impressed with the calculation.'' \\
P21 & Positive & ``The collaboration within the cultural committee was excellent... The structured selection process, the constructive exchange, and the diversity of submitted projects made the work particularly worthwhile...'' \\
P25 & Positive & ``I really have no criticism. Everyone was involved and well-informed at all times. Compliments!'' \\
\midrule
P1  & Mixed & ``Some people didn't understand that smaller projects require fewer points than larger ones. As a result, the points were sometimes distributed arbitrarily...'' \\
P11 & Mixed & ``I didn't fully understand the algorithm behind it, but I had the feeling a lot of thought went into it.'' \\
P11 & Mixed & ``No, I didn't quite understand the method—but it was good to know that somehow everything was taken into account. :-)'' \\
P19 & Mixed & ``More consideration should be given to the fact that the committee includes laypeople. The explanations sometimes assume too much; more use should be made of illustrative material...'' \\
P31 & Mixed & ``Perhaps revealing the budget later... Perhaps it could be better explained that the assessment is directly translated using the points.'' \\
\midrule
P2  & Negative & ``On decision day, clear written procedures/instructions would probably be an additional help.'' \\
P3  & Negative & ``The final voting results were also not particularly comprehensible.'' \\
P5  & Negative & ``Explain the software better. I would have liked to have better understood the process, how the projects are ultimately funded (computer program), and what weight is given to the individual evaluation and the areas of impact.'' \\
\bottomrule
\end{tabular}
\end{table}

While algorithmic explainability is often discussed in idealised terms, our field deployment highlighted the practical limitations of real-world settings. On the day of the workshop, the reveal of the Komitee Equal Shares outcome took place under significant time pressure, as it was scheduled toward the end of the event. Although participants had the opportunity to interact with the outcome table and trace their own contribution, this stage was conducted rather quickly. As a result, the explainability of the procedure—how participants could fully understand the mechanics and reasoning behind the allocation—was somewhat compromised by contextual constraints.

While the method itself allows for interaction and transparency, the time available strongly shape how well participants can make sense of it. Several participants explicitly pointed to areas where more explanation, clearer procedures, or supplementary materials would have improved their experience. Others nonetheless expressed satisfaction with the clarity of forms and the professionalism of facilitation, signalling trust in the process despite gaps in their own comprehension.

Table~\ref{tab:explainability_quotes} groups the most relevant quotes by overall sentiment: positive (satisfied and trusting), mixed (trust combined with partial confusion), and negative (requests for clearer explanations or dissatisfaction). This categorisation shows that even where comprehension was limited, trust in the fairness of the process often remained intact. 

Overall, most participants accepted the fairness of the procedure even if its details were not always fully understood. 

\subsection{Learning and Perspective Change}

As shown in Table~\ref{tab:learning_quotes}, a key outcome of participation in KK25 was the extent to which it broadened participants’ perspectives on culture, cultural workers, and the city of Winterthur. Around half of respondents (15 out of 32, 47\%) described some form of perspective change, often connecting greater awareness of the diversity and challenges of the cultural sector with a deeper appreciation of Winterthur’s cultural vibrancy.

\begin{table}[h]
\centering
\caption{Illustrative quotes on learning and perspective change.}
\label{tab:learning_quotes}
\begin{tabular}{p{0.2\linewidth} p{0.8\linewidth}}
\toprule
\textbf{Theme} & \textbf{Illustrative Quotes} \\
\midrule
\textbf{Understanding culture and cultural work}  
& ``The majority of the funds go to creative artists as wages. I found that a bit odd at first. But I realized that's exactly the point. You're doing something, and that should be rewarded. I had to let go of the idea that creative artists are just pursuing a hobby.'' (P4) \\
& ``Working on the Cultural Committee broadened my perspective on culture and its creators. I gained a better understanding of the diversity, commitment, and challenges behind cultural projects.'' (P21) \\
& ``I got an idea of how many artists depend on money and how low some of the requested amounts were.'' (P18) \\
\midrule
\textbf{Awareness of Winterthur and belonging} 
& ``I've only been living in Winterthur for a little over a year and a half. The Cultural Committee has shown me that there are so many things to do that I honestly don't take advantage of enough. (P1) \\
& ``I go through the world with more open eyes now.'' (P9) \\
& ``I'm more attentive at the moment and more interested in trying out different types of culture. I'm really looking forward to more information about the selected projects and will try to participate in many of them.'' (P16) \\
& ``I learned how many people are active in Winterthur, how creative my hometown is. The exchange with different people was an inspiration to me.'' (P19) \\
& ``My work on the Cultural Committee has shown me how diverse and engaging Winterthur's cultural scene is... This has made me even more aware of Winterthur as a vibrant and culture-friendly city.'' (P21) \\
& ``I can once again tell my friends and acquaintances that a lot is happening in the city and tell them about new projects. Winterthur is a very vibrant cultural city.'' (P25) \\
& ``The culture is diverse and interesting. I have lived in Winterthur for a long time, but I don't know everything.'' (P32) \\
\midrule
\textbf{Cross-political and social learning} 
& ``I got out of my bubble a bit. I'm on the board of the SVP Winterthur, and in my free time, I am in more conservative circles. It was good to work on a project with young, dissenting people again. Especially since I was an active Social Democrat in my youth and was involved in various youth organizations. It was very enriching.'' (P29) \\
\bottomrule
\end{tabular}
\end{table}

These reflections suggest that the Cultural Committee acted not only as a funding mechanism but also as a civic learning process. By exposing participants to the diversity, precariousness, and richness of cultural life, it fostered both a deeper appreciation of cultural work and a stronger sense of their city as a vibrant cultural community.

\section{Discussion and Conclusion}

Our study set out to examine four research questions. RQ1 asked how different allocation rules shape which projects are funded. The counterfactual analysis showed that Komitee Equal Shares (KES) leads to distinct outcomes compared to individual-only or group-only rules. It funds a broader portfolio with more, smaller projects, while maintaining close alignment with the community-defined impact fields. This reflects the method’s proportional cost sensitivity: projects succeed when both individual enthusiasm and collective values align. RQ2 examined how participants perceive their own influence, and what split between individual and group decision-making they view as fair. Self-perceptions varied widely: some felt empowered as individuals, others emphasised the strength of group deliberation. Yet across these differences, participants converged around parity—about a 50:50 balance—as the fairest design. Almost 94\% reported satisfaction with their participation opportunities, suggesting trust in the system even when perceptions of influence diverged. RQ3 asked how the Equal Shares algorithm affects behaviour and understanding. Some participants adjusted strategically while others expressed confusion, noting that more explanation would have helped. Many, however, simply trusted the algorithm’s proportional logic, reporting that it reduced the pressure to get every decision “right” and reassured them that outcomes would be fair in the aggregate. RQ4 focused on what participants learn, and how their perspective on culture and the city is reshaped. Almost half of respondents reported a change: they described new awareness of the precarious situation of cultural workers, a broadened appreciation of Winterthur’s diverse cultural landscape, and in some cases a stronger sense of civic belonging. These results show that the process is not only allocative but also educative, fostering civic learning and perspective change. Below, we distil the findings into design implications for systems that combine voting, deliberation, and algorithmic allocation.

\subsection{Design Implications}

\subsubsection{\textbf{From Aggregation to Dual Roles}}

Classical social choice has long treated voting as the simple aggregation of personal preferences: individuals reveal what they like, and a rule combines those preferences into a collective outcome. Our study, however, shows that the civic reality is more complex. Citizens do not only bring their private tastes to the table; they also learn in the process what the city needs, they develop judgments about public value, and they often act with a civic mindset that extends beyond their individual preferences. Komitee Equal Shares makes this duality explicit by distinguishing between two roles. Participants act as voters, distributing points according to their own priorities, and as evaluators, working in small impact-field groups that each spend a fixed budget proportional to the weight that the community had previously assigned to that field.  

In this sense, civic preference is not merely revealed as a static set of individual choices, but is actively formed through exposure to other perspectives, through discussion, and through the collective articulation of shared values. At the same time, this duality is not without tension. Group deliberation does more than generate evaluator signals—it also shapes individual decisions, giving people cues and guidance about which projects matter. Some participants valued this direction and saw it as a way of deepening their judgment, while others may have experienced it as too strong a nudge, making them less certain about whether their own views were being expressed independently. 

For design, this means that dual roles should not only be acknowledged but also carefully situated in context. In processes where mutual influence is seen as desirable—for example, in cultural funding, where collective values can complement individual taste—it may be appropriate to let deliberation inform personal votes. In other domains, however, it may be important to preserve independence by designing systems that keep the two roles separate, such as by preventing participants from seeing group-level judgments before they cast their own votes. Only once both sets of signals are collected can they be combined in the final allocation. Striking the right balance is therefore not a matter of principle alone, but of context: in some settings, deliberative influence can enrich civic judgment, while in others it risks undermining fairness if individuals feel their personal voice has been overshadowed.


\subsubsection{\textbf{Structuring Deliberation: Small Groups and Clear Roles}}

Deliberation quality depends as much on process design as on algorithms. In 2025, much of the workshop time was spent \emph{reading} applications rather than \emph{reasoning} about them. This suggests that the structure of group work strongly shapes the signals that deliberation produces.  

Several design choices improve focus and comparability. Groups should be small (4–6 participants) and work with limited menus (no more than 20 projects) so that projects can be discussed rather than skimmed. Criteria should be narrow and concrete (e.g., “engages youth,” “benefits new immigrants”) to anchor judgments. Unequal table budgets tied to impact-field weights should be preserved, but every ballot should be recorded as a \emph{share} (points divided by that table’s budget) so that scores are comparable across fields.  

Beyond these mechanics, facilitation, mediation, seating, and timeboxing are essential to keep discussions balanced and purposeful. When groups are small, the set of projects are focused, and roles are well defined, evaluator signals are not only easier to generate but also more reliable and useful in the final allocation.


\subsubsection{\textbf{Preparation Layers: Equipping Citizens Before Deliberation}}

When citizens are confronted with a large number of proposals, the risk is that much of the decision-making time is spent catching up on reading rather than engaging in reasoning. In Kultur Komitee 2025, participants were asked to review 30 projects online before the workshop. While this requirement encouraged early engagement, it proved insufficient: on the decision day, most participants still encountered many new and unfamiliar projects, and substantial time slipped into basic reading rather than discussion.  

This highlights the need for a stronger preparation layer to ensure that participants arrive better equipped and confident in their roles. Such preparation can be embedded in the process design itself. Light but structured pre-work—such as ensuring every project is seen by multiple reviewers or distributing thematic overviews—can normalise arriving informed and help people participate on more equal terms.  

For future direction, technical tools can complement this civic practice. Interactive visualisations could provide an accessible overview of the proposal landscape, highlighting clusters, budget sizes, or links to impact fields. Recommender systems could guide each participant to explore a diverse set of projects in advance, balancing coverage with relevance and surfacing areas where they may bring expertise. Large language model assistants could further reduce the burden by answering targeted questions (for example, “Which projects focus on public space for youth?” or “Which proposals are most affordable within the tradition field?”) and by offering succinct comparisons.  

The design implication is that preparation should be treated as an integral part of the participatory process rather than an optional add-on. By combining structured pre-work with supportive tools, committees can shift deliberation time from catch-up to informed reasoning, making the collective signals they produce more reliable and meaningful.



\subsubsection{\textbf{Priceability as Explainability and an ``Open Review'' Model for Grantmaking}}

Priceability can be framed as a requirement of explainability in civic allocation \citep{Rey2023}. A project should only be funded when the combined budgets of its supporters can cover its cost, and each decision must be expressible as an itemised payment. In Komitee Equal Shares, this principle was extended into a tangible feature: voting \emph{receipts} that show how individual votes and impact-field budgets jointly contributed to each funded project. Such receipts turn justification into an artefact that participants can inspect, reproduce, and contest. In contrast to opaque algorithmic systems, priceability makes collective decisions inherently traceable.  

Under the right conditions, this logic could support an \emph{open review} model for grantmaking. Funded sets could be published together with per-project receipts and plain-language explanations of the rules and their assumptions. These artefacts should make clear not only which projects were selected, but also why others were skipped, for example, because projects within the same field exhausted its budget. In this way, participants and the wider public can trace the outcome in terms they understand, linking numerical receipts to qualitative reasoning. In this ideal model, these materials should be both human-readable and machine-readable, enabling journalists, community groups, and researchers to audit, recompute, and contest results. However, at the same time, any open review model must be embedded in context. In real-world settings, issues of voting privacy, political sensitivity, and the risk of misuse all matter. Open review is therefore best understood a design ideal that can improve accountability, but one that must be adapted to local conditions to balance transparency with protection and trust.

\subsection{Limitations and Future Work}

This study is based on a single case of cultural grantmaking in Winterthur. While the Kultur Komitee is a rich and innovative example, cultural funding has particular characteristics—modest budgets, project-based applications, and a mix of professional and grassroots actors—that may not generalise to other policy areas or higher-stakes settings. Future research should test Komitee Equal Shares in different contexts to examine its robustness and adaptability. The scale and composition of the committee also pose limits. Each year around 30–40 citizens take part, drawn by lot but without demographic quotas. This ensures diversity but not representativeness, and it restricts the statistical power of survey analysis. Larger or comparative deployments could provide stronger evidence on how participants perceive fairness, influence, and role clarity.  

The process itself required considerable engagement, asking participants to deliberate, vote, and assign weights to impact fields. While this was feasible in Winterthur, the time and cognitive load may be barriers in other settings. Future work should explore simplified designs or AI-supported methods that reduce cognitive burden while preserving agency.  

Another limitation concerns explainability in practice. Although the priceable allocation rule provides receipts, the communication of why projects were funded or skipped was constrained by time and format. Some participants remained unsure how to interpret outcomes. Research should test alternative modes of explanation—such as interactive receipts, simulations, or narrative visualisations—and evaluate their effects on trust and learning. Also, transparency is valuable, but in practice it must be balanced with voting privacy, political sensitivities, and the risk of misuse. Future studies could investigate how different publics want transparency to be implemented and how openness can coexist with protections for participants.  

Finally, algorithmic and temporal factors deserve more attention. Komitee Equal Shares extends the Method of Equal Shares, but other proportional rules or different tie-breaking strategies could produce different portfolios and different perceptions of fairness. Comparative studies would help to clarify these trade-offs. Long-term effects also remain unknown. Our evaluation captured immediate perceptions and learning, but not whether participation changes cultural engagement, trust in institutions, or attitudes toward public spending over time. Longitudinal research will be necessary to address these questions.

\section{Conclusion}

The Komitee Equal Shares framework shows how voting and deliberation can be combined into a single proportional, priceable allocation that is both fair and explainable. In the Kultur Komitee Winterthur, the approach produced balanced portfolios, clarified participants’ dual roles, and fostered civic learning, while also revealing challenges of preparation and explainability under real-world constraints. The framework contributes to participatory grant-making and budgeting by reframing participation as both preference expression and public evaluation, and by extending the Method of Equal Shares into a priceable rule that generates voting receipts. These receipts turn justification into a built-in artefact, making collective choices traceable and contestable. For design, the lesson is that participatory systems should model roles explicitly, embed traceability directly in the allocation rule, and structure deliberation and equip citizens to avoid information overload. More broadly, Komitee Equal Shares demonstrates how algorithmic rules can support not only fair allocation, but also transparent and trustworthy decision-making in civic funding.

\subsection{Acknowledgements}
This research was made possible through the collaboration and support of many individuals and institutions. The Komitee Equal Shares approach was implemented in close cooperation and sustained dialogue with Mia Odermatt, co-project lead of Kultur Komitee (SKKG). We especially acknowledge Rahel Stauffiger (SKKG), whose expertise in process design and evaluation was central to the qualitative data collection. The Kultur Komitee and SKKG are gratefully recognised for their openness to innovation and experimentation, which created the institutional setting for this study. We thank the participants and project owners of the Kultur Komitee for their active engagement, which provided the foundation for this scientific endeavour. We also acknowledge the contributions of Fynn Bachmann (UZH) for his assistance in carrying out the process and his reflections on deliberative practices, as well as Andreas Geis for his support in organisational strategy. Our thanks further extend to The Fundament for enabling the digital application used to organise the projects and to support the pre-workshop learning phase. Finally, we express our gratitude to Professor Dirk Helbing (ETH Zurich) and Dr. Carina I. Hausladen (ETH Zurich) for their intellectual guidance and encouragement throughout the development of this work.

\bibliographystyle{ACM-Reference-Format}
\bibliography{main}

\end{document}